
\input phyzzx

\def\np{Nucl. Phys.}
\def\pl{Phys. Lett.}

\def\cmp{Comm. Math. Phys.}
\def\ijmp{Int. J. Mod. Phys.}
\def\mpl{Mod. Phys. Lett.}

\def\am{Ann. of Math.}

\def\phyrep{Phys. Rep.}
\def\am{Adv, in Math.}

\tolerance=500000
\overfullrule=0pt
\Pubnum={US-FT-3/93}
\pubnum={US-FT-3/93}
\date={June, 1993}
\pubtype={}
\titlepage

\title{$gl(N,N)$ CURRENT ALGEBRAS AND TOPOLOGICAL FIELD
THEORIES} \author{J. M. Isidro and A. V. Ramallo}
\address{Departamento de F\'\i sica de Part\'\i culas \break
Universidad de Santiago \break
E-15706 Santiago de Compostela, Spain}

\abstract{The  conformal field theory for the $gl(N,N)$
affine Lie superalgebra in two space-time dimensions is
studied. The energy-momentum tensor of the model, with
vanishing Virasoro anomaly, is constructed. This theory has a
topological symmetry  generated by
operators of dimensions 1, 2 and 3, which are represented as
normal-ordered products of $gl(N,N)$ currents. The topological
algebra they satisfy is linear and differs from the one obtained
by twisting the $N=2$ superconformal models. It closes with a set
of $gl(N)$ bosonic and fermionic currents. The
Wess-Zumino-Witten model for the supergroup $GL(N,N)$ provides
an explicit realization of this symmetry and can be used to
obtain a free-field representation of the different
generators. In this free-field representation, the theory
decomposes into two uncoupled components with $sl(N)$ and $U(1)$
symmetries. The non-abelian component is responsible for
the extended character of the topological algebra, and it is
shown to be equivalent to an $SL(N)/ SL(N)$ coset model.
In the light of these results, the $G/ G$ coset models
are interpreted as topological sigma models for the group
manifold of $G$.}

\endpage
\pagenumber=1

\chapter{Introduction and summary of main results}

Since their  introduction by Witten
\REF\wittop{E. Witten \journal\cmp&117(88)353.}
[\wittop] five years ago, topological field theories have been a
subject of intensive  investigation . Some of these theories
have been shown to be
relevant in the attempts to understand the
non-perturbative structure of string
theory and quantum gravity\REF\witgrav{E. Witten
\journal\np&B340(90)281.}[\witgrav].
Moreover the
three-dimensional
Chern-Simons gauge theory
\REF\witCS{E. Witten \journal\cmp&121(89)351.}[\witCS]
has provided us with a
fascinating connection between
three-dimensional topology and conformal field theory in two
space-time dimensions.

In Mathematics,  topological field theories have become
valuable tools in the study of topological invariants of
low-dimensional manifolds. So, for example, the Donaldson
polynomials are obtained from observables of the
four-dimensional topological Yang-Mills theories [\wittop], and
two-dimensional topological gravity
\REF\LPW{J.M.F. Labastida, M. Pernici and E.
Witten\journal\np&B310(88)611.}[\LPW] provides a framework to
study intersection theory on the moduli space of Riemann
surfaces
\REF\konse{M. Konsevich \journal\cmp&147(93)1.}
[\witgrav,\konse]. In three dimensions, Chern-Simons theories
allow to define invariant polynomials for knots and links and,
for example, when the gauge group is $SU(2)$, the Jones
polynomial and its generalizations are obtained.

In general there are two fundamental classes of topological
theories
\REF\bbrt{For a review see D. Birmingham, M. Blau, M.Rakowski
and G. Thompson \journal\phyrep&209(91)129.}[\bbrt].
To the first class  belong those
models in which the action does not depend on the metric of
the manifold on which the local fields are defined
\REF\horo{G.T. Horowitz \journal\cmp&125(89)417;
 G.T. Horowitz and M. Srednicki
\journal\cmp&130(90)83.}[\horo]. The natural
observables of these models are metric-independent operators
whose vacuum expectation values give rise to topological
invariant quantities. Examples of this first class of theories
are the $BF$ models and the already mentioned Chern-Simons
theory.

The second class of topological theories have an action which
does depend on the metric of the base manifold. However, these
theories possess a nilpotent fermionic symmetry that allows the
introduction of a BRST cohomology. The distinctive feature of
this class of theories is that their energy-momentum tensor
is exact within this BRST cohomology, a fact which ensures the
metric-independence of any correlator involving BRST-invariant
operators. This is precisely the sense in which these
theories are topological. The Donaldson-Witten theory in four
dimensions [\wittop], and the topological sigma model in two
\REF\wit{E. Witten \journal\cmp&118(88)411.} [\wit],
are the most outstanding theories of this class. In this paper
we shall  consider  this second type of topological field
theories.

Several procedures have been proposed in order to construct
models satisfying the highly non-trivial constraint of having
an energy-momentum tensor which is exact with respect to a
fermionic symmetry of the action. The mostly used one is that
 in which the topological theory is obtained from models
having two or more supersymmetries. This procedure is based on
a redefinition of the Lorentz group of the theory, which
amounts to a modification (a ``twist") of the energy-momentum
tensor $T$ of the model in such a way that it becomes the BRST
variation of a fermionic operator
\REF\EY{T. Eguchi and S.-K. Yang \journal\mpl&A4(90)1653;
T. Eguchi, S. Hosono and S.-K. Yang \journal\cmp&140(91)159.}
[\EY]. The BRST current
implementing the topological symmetry of the twisted theory is
one of the supercurrents of the model we start with.
Applying this twist to an $N=2$ superconformal theory in
two dimensions
\REF\LVW{W. Lerche, C. Vafa and N.P. Warner
\journal\np&B324(89)427.}[\LVW], one ends up with a topological
conformal field theory
\REF\dij{R. Dijkgraaf, E. Verlinde and H. Verlinde
\journal\np&B352(91)59.;``Notes on topological string theory
and 2d quantum gravity", Proceedings of the Trieste spring
school 1990, edited by M.Green et al. (World Scientific,
Singapore,1991).}[\dij]. The generators of this topological
symmetry close an algebra (the so-called conformal topological
algebra) which is a transcription of the original $N=2$
superconformal algebra. The main drawback of this twisting
procedure is the difficulty of constructing models possessing
more than one supersymmetry. Furthermore, it has been shown
\REF\llatas{J.M.F. Labastida and P.M.
Llatas\journal\np&379(92)220.}
[\llatas ] that one can relax the conditions required to the
untwisted theory while keeping a sensible topological theory
after the twist is performed.

Another method to generate topological conformal theories in
two dimensions has been proposed by Eguchi and Yang [\EY]. It
consists in constructing coset theories with vanishing Virasoro
central charge $c$. These cosets are formulated in terms of
bosonic operators. However, at the topological point $c=0$, a
fermionic symmetry making $T$ BRST-exact shows up. In a Coulomb
gas approach, the generator of this topological symmetry is
obtained from the screening operators of the theory. Some other
procedures have been presented. In general, the requirements
that a topological field theory must fulfil make their
construction so restrictive that one may hope that a complete
classification programme of (at least) some classes of them
could be completed.

In this paper a new method to generate topological
conformal field theories is presented. Our construction makes
use of the properties of  affine Lie superalgebras
\REF\kac{V. Kac \journal\am&26(77)8;
V. Kac \journal\cmp&53(77)31.}
\REF\scheu{M. Scheunert, `` The theory of Lie superalgebras",
Springer-Verlag, Berlin 1978.}[\kac,\scheu] whose bosonic
and fermionic contents are matched in such a way that the net
balance of commuting and anticommuting local degrees of
freedom of the corresponding conformal field theory gives a
vanishing result. In order to implement this balance we will
be forced to consider non-semisimple Lie superalgebras
and, in fact, we shall restrict ourselves to the case of
$gl(N,N)$.

Our first step
will be the construction of the energy-momentum tensor $T$ of
the theory. This operator  will be obtained by requiring
the $gl(N,N)$ currents to be primary, dimension-one operators.
The fact that we are dealing with a non-semisimple Lie algebra
 will introduce additional complications
to the standard Sugawara construction of $T$. These new features
were analysed by Rozasnki and Saleur
\REF\RS{L. Rozanski and H. Saleur \journal\np&B376(92)461;
Yale preprint YCTP-P10-92.}[\RS], who considered the
$gl(1,1)$ case in their study of the Alexander-Conway knot
polynomial in quantum field theory
\REF\KS{L. Kauffman and H. Saleur
\journal\cmp&141(91)293.}[\KS]. Their method carries over
to the more general $gl(N,N)$ model. Once $T$ has been
determined, we shall check that the corresponding conformal
field theory has a vanishing Virasoro anomaly, and we will
characterize its underlying topological symmetry. In our
approach there is a manifest balance between even and
odd excitations, and it is straightforward to find the
symmetry  relating them. Actually, this symmetry can be
generated in a local way by a suitable combination of the
fermionic currents of the affine Lie superalgebra. The
exactness of our energy-momentum tensor with respect to this
topological symmetry can be readily verified. The current
algebra is used as a guiding principle for this purpose and,
as a matter of fact, the odd partner of $T$ is constructed by a
fermionic analogue of the Sugawara construction. In
general, all the operators appearing in the topological algebra
will be normal-ordered products of currents.

The elucidation of
the algebra closed by $T$ and its odd companion can serve to
unravel the algebraic structures underlying the
topological symmetry in quantum field theory. The
algebra of our model transcends the mere twisting of the $N=2$
superconformal algebra. In general, we shall obtain an
extended topological algebra  including two additional
dimension-three operators (a bosonic field and its fermionic
partner). An interesting aspect of this algebra is its linear
character, in spite of the fact that it contains higher-spin
operators. Actually, Kazama
\REF\kaza{Y. Kazama \journal\mpl&A6(91)1321.}[\kaza]
found this same algebra as a consistent truncation of a
larger one, satisfied by the matter system studied by Distler
\REF\dist{J. Distler \journal\np&B342(90)523.}[\dist].
Therefore our model provides an explicit realization of
this extended topological algebra.

It is important to point out that the model we shall
construct is already ``twisted" , \ie, no redefinition  is
needed  to render the theory topological. Nevertheless, we can
try to relate our algebra to the standard superconformal
symmetries. It turns out that, upon ``untwisting", our algebra
can be related to the $N=1$ superconformal algebra in such a way
that all generators can be arranged into $N=1$ supermultiplets.

Another topic we shall investigate is the compatibility of
the extended topological algebra with the current algebra on
which it is based. By construction, $T$ closes with the
currents when they are commuted, so it is
quite natural to require  the latter to be primary also with
respect to the odd partner of $T$. It turns out, however, that
only half the initial currents satisfy this condition.
They close with themselves and
with the generators of the topological algebra. Each of the
bosonic currents in this restricted set is
BRST-exact, and their fermionic partners also belong to this
set. In fact, they close an algebra that is the topological
analogue of the affine algebra and that we shall name
topological current algebra.  Our analysis will reveal that
the topological symmetry can coexist with a current algebra.
Actually, the non-abelian nature of the latter determines the
extended character of the former. As  happens with
conformal field theories, one would expect that additional
symmetries, such as current algebra symmetries, could serve to
organize the Hilbert space of the topological theory in such a
way that a well-defined representation theory could be
developed. In string theory this extra symmetry could provide a
dynamical principle to overcome the difficulties appearing
beyond the $d=1$ barrier
\REF\gins{For a review see P. Ginsparg and G. Moore,
``Lectures on 2D Gravity and  2D String
Theory"(hep-th/9304011), Yale preprint YCTP-P23-92.}[\gins].

An explicit realization of this topological conformal field
theory can be obtained by quantizing the
Wess-Zumino-Witten (WZW) model
\REF\WZW{E. Witten \journal\cmp&92(84)455.}[\WZW] for the
$GL(N,N)$ supergroup. Using this realization we shall conclude
that our model describes a zero-dimensional topological sigma
model. Performing a Gauss-type decomposition of the  of the
basic $GL(N,N)$ variable of the WZW model, we shall obtain a
free-field representation
\REF\gera{A. Gerasimov et al. \journal\ijmp&A5(90)2495.}[\gera]
 of the extended topological algebra
that will shed light on its nature and will clarify its
relation with other models of two-dimensional topological
matter. Within this free-field realization, our model is
represented as the superposition of two uncoupled models
having $sl(N)$ and $U(1)$ symmetries. Each of these two
 separately constitutes a topological conformal field theory.
The non-abelian one realizes  the extended
topological algebra non-trivially, whereas for the $U(1)$
component, this algebra reduces to the one obtained by twisting
the $N=2$ superconformal models. The non-abelian theory we
shall find is identical to the representation of the $G/G$ coset
models
\REF\witGG{E. Witten \journal\cmp&144(92)189.}
[\witGG] obtained in references
\REF\yank{M. Spiegelglas and S. Yankielowicz
\journal\np&393(93)301.}
\REF\aharo{O. Aharony et al. Tel Aviv University preprint,
TAUP-1961-92 \journal\pl&B289(92)309 \journal\pl&B305(93)35.}
\REF\hu{H.L. Hu and M. Yu \journal\pl&B289(92)302
\journal\np&B391(93)389.} [\yank,\aharo,\hu]. This means
that our extended topological algebra is realized in this type
of topological theories which, from our point of view, are
regarded as topological sigma models for group manifolds, \ie,
as the topological analogue of the Wess-Zumino-Witten
theories. These aspects of our construction will be discussed
elsewhere
\REF\letter{J.M. Isidro and A. V. Ramallo, ``Topological
current algebras in two dimensions", Santiago preprint, to
appear.} [\letter].

This paper is organized as follows. In section 2, after
reviewing the basic features of the $gl(N,N)$ current
algebras needed in this paper, we construct the
energy-momentum tensor. The
topological algebra of our model is explored in section 3.
The relations satisfied by this algebra have been compiled in
 Appendix A. The topological current algebra
of our model is  obtained in  section 4. An example of
how these topological algebras appear in the $sl(2)$
WZW model is developed in Appendix B. In section 5 the
free-field representation of the extended topological algebra
is worked out. Finally, in section 6 we discuss our results
and indicate some possible lines of future development of our
ideas.

\chapter{$gl(N,N)$ current algebras}

We begin this section  giving the basic definitions
and properties of the $gl(N,N)$ Lie superalgebra [\kac,\scheu ].
Consider a supervector space having $N$ bosonic dimensions and
$N$ fermionic ones. In this supervector space we shall take an
homogeneous basis having $N$ bosonic and $N$ fermionic elements.
We shall label this basis in such a way that the first $N$
vectors are the bosonic ones. Accordingly, in this space, any
vector is determined by a set of quantities  $V^A$ with $1\leq A
\leq 2N$,  the $V^a$($V^{a+N}$) for $a=1, \cdots , N$ being the
bosonic (fermionic) components. In what follows a capital
latin letter will denote an index running from $1$ to $2N$,
whereas lower-case latin indices can take any value
between $1$ and $N$. The grade of a given index $A$
(denoted by $g(A)$) is defined to be zero (one) if it labels
a bosonic (fermionic) component. Therefore, with our
conventions,
$$
g(A)=\cases{0, &if $1\le A\le N$\cr
            1, &if $N+1\le A \le 2N$.\cr}
\eqn\uno
$$
The $gl(N,N)$ Lie superalgebra is the algebra of $2N\times
2N$ matrices acting on a supervector space with $N$
bosonic and $N$ fermionic dimensions. In an homogeneous
basis we can write the general form of any element of
$gl(N,N)$ as
$$
X=\pmatrix{B_{1} & F_{1}\cr F_{2}  & B_{2}\cr},
\eqn\dos
$$
where $B_{1}$ and $B_2$ are $N\times N$ matrices whose
elements are c-numbers, whereas the entries of the $N\times
N$ matrices $F_1$ and $F_2$ are odd Grassmann numbers. The
matrices of the form
$$
B=\pmatrix{B_{1} & 0\cr 0 & B_{2}\cr}
\,\,\,\,\,\,\,\,\,\,\,\,
F=\pmatrix{0 & F_{1}\cr F_{2} &0\cr},
\eqn\tres
$$
are said to be homogeneous. To these homogeneous matrices
we shall assign a grade. A matrix like $B$ is called
even, whereas $F$ is said to be odd. Actually we shall
consider off-diagonal matrices like $F$ as odd matrices
even if their entries are c-numbers. The grade of any
homogeneous matrix X is defined as follows:
$$
g(X)=\cases{0, &if X is even\cr
            1, &if X is odd. \cr}
\eqn\cuatro
$$
The grade of vector indices and matrices (eqs. \uno and
\cuatro ) are defined modulo $2$ (\ie\ as for a $Z_2$
grading).

The superalgebra structure for $gl(N,N)$ is introduced by
defining the generalized Lie bracket for any two
homogeneous matrices $X$ and $Y$:
$$
[X,Y] \equiv XY- (-1)^{g(X)g(Y)}YX.
\eqn\cinco
$$
Notice that if at least one of the two matrices $X$ and
$Y$ is even, their  bracket $[X,Y]$ is a commutator while,
on the contrary, if both $X$ and $Y$ are odd, $[X,Y]$ is an
anticommutator. Using the definition of the bracket given
above, it is easy to check that it satisfies a graded
Jacobi identity:
$$
\eqalign{(-1)^{g(X_1)g(X_3)}[X_1,[X_2,X_3]]+&
(-1)^{g(X_2)g(X_1)}[X_2,[X_3,X_1]]+\cr+&
(-1)^{g(X_3)g(X_2)}[X_3,[X_1,X_2]]=0.\cr}
\eqn\seis
$$
Furthermore, the Lie bracket \cinco\ acts as a graded
derivation, \ie\ it satisfies the identity
$$
[X_1X_2,X_3]=(-1)^{g(X_2)g(X_3)}[X_1,X_3]X_2+
X_1[X_2,X_3].
\eqn\siete
$$
Let us now define the matrices $E_{AB}$ as follows:
$$
(E_{AB})_{CD}=\delta_{AC}\delta_{BD}.
\eqn\ocho
$$
Obviously the set $\lbrace E_{AB} ; 1\leq A,B\leq 2N
\rbrace $ constitute a basis for the space of $2N\times
2N$ matrices. Therefore any $X\in gl(N,N)$ can be written
as
$$
X=\sum_{A,B=1}^{2N}X^{AB}E_{AB},
\eqn\nueve
$$
where the $X^{AB}$ are the contravariant components of $X$
with respect to the basis $\lbrace E_{AB} \rbrace $.  From
our previous definitions, the grade of a matrix $E_{AB}$
is given by
$$
g(E_{AB})=g(A)+g(B).
\eqn\diez
$$
Notice that the $E_{AB}$ matrices with zero grade span the
c-number part of $gl(N,N)$, whereas those with
non-vanishing grade span (with odd Grassmann coefficients)
the odd part of $gl(N,N)$. It is interesting to compute
the generalized Lie brackets among  the $E_{AB}$
matrices. Using their explicit form in the definition of
the bracket (eq. \cinco) it is straightforward to arrive
at
$$
[E_{AB},E_{CD}]=\sum_{P,Q=1}^{2N}
F_{AB,CD}^{PQ}E_{PQ},
\eqn\once
$$
where we have written the result in terms of the
structure constants
$$
F_{AB,CD}^{PQ}=\delta_{AP}\delta_{DQ}\delta_{BC}-
(-1)^{(g(A)+g(B))(g(C)+g(D))}
\delta_{CP}\delta_{BQ}\delta_{AD}.
\eqn\doce
$$
Although we have obtained the Lie brackets \once\ from
the explicit expression for the matrix elements  of $E_{AB}$
(eq. \ocho), we can now regard eqs. \once\ and \doce\ as the
definition of the abstract Lie superalgebra $gl(N,N)$. The
$E_{AB}$ are the generators of this algebra, while their
explicit form written down in eq. \ocho\ constitutes the
defining (fundamental) representation of $gl(N,N)$. From
eq. \doce\ we easily obtain the (anti)symmetry properties
of the structure constants
$$
\eqalign{
F_{CD,AB}^{PQ}=&
-(-1)^{(g(A)+g(B))(g(C)+g(D))}F_{AB,CD}^{PQ}\cr
F_{AB,CD}^{PQ}=&-(-1)^{g(Q)+g(A)}
(-1)^{(g(A)+g(B))(g(P)+g(Q))}F_{QP,CD}^{BA}.\cr}
\eqn\catorce
$$
These equations will be very useful in future
calculations. For a matrix like \dos\ one can define the
supertrace as follows:
$$
Str(X)=Tr(B_1)-Tr(B_2)=\sum_{A=1}^{2N}(-1)^{g(A)}X^{AA},
\eqn\quince
$$
where $X^{AA}$ are the diagonal contravariant components
of $X$ with respect to the $\lbrace E_{AB} \rbrace $ basis
(see eq. \nueve). The supertrace \quince\ can be used to
define the Killing-Cartan bilinear form for $gl(N,N)$. In
complete analogy with what it is done for ordinary Lie
algebras, we define the following invariant bilinear form:
$$
<X,Y>\equiv Str(XY),
\eqn\dseis
$$
where $X$ and $Y$ are two matrices in the fundamental
representation of $gl(N,N)$. The inner products of the
elements of the $\lbrace E_{AB} \rbrace $ basis define the
metric tensor:
$$
G_{AB,CD}\equiv <E_{AB},E_{CD}>=Str(E_{AB}E_{CD}).
\eqn\dsiete
$$
Using the explicit expressions of the matrix elements of
the $E_{AB}$ matrices, one immediately gets
$$
G_{AB,CD}=(-1)^{g(A)}\delta_{AD}\delta_{BC}.
\eqn\docho
$$
Notice that, contrary to what happens for ordinary Lie
algebras, $G_{AB,CD}$ is not symmetric. This has to be taken
into account when performing explicit calculations.
The inverse matrix of $G_{AB,CD}$ will be denoted by
superindices. It satisfies
$$
\sum_{C,D} G_{AB,CD}G^{CD,PQ}=
\sum_{C,D} G^{AB,CD}G_{CD,PQ}=\delta_{AP}\delta_{BQ}.
\eqn\dnueve
$$
These two conditions are fulfilled by

$$
G^{AB,CD}=(-1)^{g(B)}\delta_{AD}\delta_{BC}.
\eqn\veinte
$$
Using the inverse metric tensor, we can obtain the
quadratic Casimir operator of the $gl(N,N)$ algebra:
$$
C_1=\sum_{A,B,C,D}G^{AB,CD}E_{AB}E_{CD}=
\sum_{A,B} (-1)^{g(B)}E_{AB}E_{BA}.
\eqn\vuno
$$
{}From the fundamental Lie brackets \once\ and the
derivation property \siete\ one can prove, after a short
calculation, that the bracket of $C_1$ with any element of
the algebra vanishes:
$$
[C_1,E_{AB}]=0.
\eqn\vdos
$$
It is important to point out that \vdos\ can be proved
without using the explicit expressions of the $E_{AB}$
matrices. Therefore \vdos\ is valid for any representation
of the abstract Lie superalgebra. For the fundamental
representation \ocho\ $C_1$ actually vanishes:
$$
(C_1)_{AB}^{Fundamental}=0.
\eqn\vtres
$$
The $gl(N,N)$ superalgebra is not semisimple. It contains the
element $\sum_A E_{AA}$, which has a vanishing bracket  with
all the generators (in the fundamental representation this
element is represented by the identity matrix). Due to this
fact we have a second quadratic Casimir:
$$
C_2=\sum_{A,B} E_{AA}E_{BB}.
\eqn\vcuatro
$$
Again one easily proves using  \siete\ that
$$
[C_2,E_{AB}]=0.
\eqn\vcinco
$$
For the fundamental representation $C_2$ is just the unit
matrix,
$$
(C_2)_{AB}^{Fundamental}=\delta_{AB}.
\eqn\vseis
$$

Let us consider now the affine  Kac-Moody
superalgebra based on $gl(N,N)$. Our basic object will be an
holomorphic current $J(z)$ taking values in a $gl(N,N)$
algebra. The contravariant components $J^{AB}(z)$ of the
currents are the coefficients of the expansion of $J(z)$ in
terms of the $\lbrace E_{AB} \rbrace $ basis:
$$
J(z)=\sum_{A,B} J^{AB}(z) E_{AB}.
\eqn\vsiete
$$
Notice that, since $J(z)$ takes values in $gl(N,N)$, the
components $J^{a,b}(z)$ and $J^{a+N,b+N}(z)$ are bosonic
(\ie\ c-number valued) and, on the contrary, $J^{a,b+N}(z)$
and $J^{a+N,b}(z)$ are fermionic ($1\leq a,b \leq N $).
The covariant components $J_{AB}(z)$ are given by
$$
J_{AB}(z)=Str(J(z)E_{AB})=(-1)^{g(B)}J^{BA}(z),
\eqn\vocho
$$
where in the last step we have made use of the explicit
form of the metric tensor (see eq. \docho). In order to
define the affine Kac-Moody superalgebra it is convenient
to expand $J_{AB}(z)$ in a Laurent series around $z=0$:
$$
J_{AB}(z)=\sum_{n\in Z}J_{AB}^n z^{-n-1}.
\eqn\vnueve
$$
The $gl(N,N)$ current algebra is obtained by requiring
 the modes $J_{AB}^n$ to satisfy
$$
[J_{AB}^n,J_{CD}^m]= F_{AB,CD}^{PQ}J_{PQ}^{n+m}
+kn\delta_{n+m,0}G_{AB,CD},
\eqn\treinta
$$
where $F_{AB,CD}^{PQ}$ are the structure constants \doce\
and the central extension has been taken to be proportional
to  the metric tensor $G_{AB,CD}$. In \treinta\ $k$ is
a c-number constant (the level of the algebra). An
alternative way of defining the current algebra is
obtained by giving the short-distance expansion of the
product of two arbitrary currents:
$$
J_{AB}(z)J_{CD}(w)={k\over
(z-w)^2}G_{AB,CD}+F_{AB,CD}^{PQ}{J_{PQ}(w)\over z-w}.
\eqn\tuno
$$
As is well known in the framework of the radial
quantization of  two-dimensional field theories, the
operator product expansions (OPE) \tuno\ are equivalent to
the brackets \treinta. We shall  indistinctly use OPE's or
brackets as best suits our convenience.

We can undertake now the
construction of a two-dimensional conformal invariant
theory consistent with the $gl(N,N)$ affine
symmetry. This objective would be accomplished if we
were able to construct an energy-momentum tensor $T$
such that the currents $J_{AB}(z)$ transform as dimension-one
primary fields. The natural ansatz for $T$ is the Sugawara
construction, in which $T$ is built up as a quadratic
expression in the currents. In order to unambiguously define
these operators, in which two or more fields evaluated
at the same point are multiplied, we need to adopt a normal
ordering prescription. Suppose that $A(z)$ and $B(z)$ are two
local fields whose Laurent modes are $A^n$ and $B^n$,
$$
A(z)=\sum_{n\in Z}A^n z^{-n-\Delta_A}
\,\,\,\,\,\,\,\,\,\,\,\,\,\,
B(z)=\sum_{n\in Z}B^n z^{-n-\Delta_B},
\eqn\tdos
$$
where $\Delta_A$ and $\Delta_B$ are the conformal weights
of $A(z)$ and $B(z)$ respectively. All the fields we shall
encounter inside normal-ordered products will have integer
conformal weights and, therefore, we shall assume that this
condition is satisfied in the equations that follow.
The normal-ordered
product of two arbitrary modes $:A^n B^m:$ is defined as
$$
 :A^n B^m:\,\,\equiv
\cases{A^n B^m, &if $m \geq 1-\Delta_B $\cr
       (-1)^{g(A)g(B)} B^m A^n   , &if $m < 1-\Delta_B$.\cr}
\eqn\ttres
$$
The modes $(:AB:)^n$ of the normal-ordered product of
$A$ and $B$ are defined by the equation
$$
:A(z)B(z): \,\, \equiv
\sum_{n\in Z} (:AB:)^n z^{-n-\Delta_A -\Delta_B}.
\eqn\tcuatro
$$
Substituting the mode expansions of $A(z)$ and $B(z)$ in
the left-hand side of \tcuatro\ and using \ttres\ we get
$$
(:AB:)^n= \sum_{p=1-\Delta_B}^{\infty} A^{n-p} B^p +
 (-1)^{g(A)g(B)}\sum_{p=\Delta_B}^{\infty}B^{-p}A^{n+p}.
\eqn\tcinco
$$
It is important to point out that the order of
fields inside a normal-ordered product is relevant. Indeed, one
has $$
(:AB:)^n=(-1)^{g(A)g(B)}(:BA:)^n+
\sum_{p=1-\Delta_B}^{n+\Delta_A-1} [A^{n-p}, B^p].
\eqn\tseis
$$
Notice that, apart from a sign, we get an extra
contribution when we reverse the order of the
operators $A$ and $B$ inside $:AB:$. Let us examine the
consequences of eq. \tseis\ when $A$ and $B$ have a bracket of
the form
$$
[A^n,B^m]=D^{n+m}+nk_D\delta_{n+m,0},
\eqn\tseisi
$$
where $D$ is an operator of conformal dimension $\Delta_D$,
which
is easily seen to be $\Delta_A+\Delta_B-1$, and $k_D$ is a
constant number. Using \tseisi\ in eq. \tseis, we get
$$
(:AB:)^n=(-1)^{g(A)g(B)}(:BA:)^n-(\partial D)^n
-{\Delta_D\over 2}(\Delta_A-\Delta_B)k_D\delta_{n,0}
\eqn\tseisii
$$
In eq. \tseisii\ we have introduced the derivative of the
operator $D$, whose modes are given by
$$
(\partial D)^n=-(n+\Delta_D)D^n.
\eqn\tseisiii
$$
Notice that when $\Delta_A=\Delta_B$ (which will
be the case in
most of our calculations), the last term in
the right-hand side of
\tseisii\ disappears and only the term containing
$\partial D$ survives.

When more than two fields are multiplied, the
normal order is defined inductively according to
the rule
$$
:A_n\cdots A_1:\equiv
:(:(:\cdots (:A_nA_{n-1}:)\cdots :)A_2 :)A_1):,
\eqn\tsiete
$$
\ie, the product $:A_n\cdots A_1:$ is considered as the
product of $:A_n\cdots A_2:$ with $A_1$ and so on. A
reordering formula like \tseisii\ can also be obtained for
normal-ordered products of more than two fields. Proceeding as
we did to get eq. \tseisii\ and using the prescription
\tsiete\ one obtains
$$
\eqalign{
(:ABC:)^n=&(-1)^{g(A)g(B)}(:BAC:)^n-(\partial D C)^n
-{\Delta_D\over 2}(\Delta_A-\Delta_B)k_D C^n\cr
(:CAB:)^n=&(-1)^{g(A)g(B)}(:CBA:)^n-(C\partial D )^n
-{\Delta_D\over 2}(\Delta_A-\Delta_B)C^n k_D,\cr}
\eqn\tsietei
$$
where we have supposed that the bracket \tseisi\ still holds.

The  obvious candidates to become the
energy-momentum tensor of the theory are the Sugawara
bilinears constructed from the quadratic Casimir
invariants of the theory. As we have seen, the $gl(N,N)$
algebra has two such Casimir invariants and,
therefore, we should consider the operators
$$
\eqalign{
T_1=&\sum_{A,B}(-1)^{g(B)}:J_{AB}J_{BA}:\cr
T_2=&\sum_{A,B}:J_{AA}J_{BB}:.\cr}
\eqn\tocho
$$

In order to determine the precise combination of $T_1$ and
$T_2$ that makes $J_{AB}$ a primary field with conformal
weight equal to one, we must compute the brackets
$[T_1^n,J_{AB}^m]$ and $[T_2^n,J_{AB}^m]$. These brackets
can be obtained from a general expression that we shall now
derive. Suppose  we want to compute
$$
[ (:A_1A_2:)^n, A_3^m],
\eqn\tnueve
$$
where $A_1$, $A_2$ and $A_3$ are operators of dimensions
$\Delta_1$, $\Delta_2$ and $\Delta_3$ whose
Laurent modes have the
brackets  $$
\eqalign{
[A_1^n,A_3^m]=&A_{13}^{n+m}+nk_{13}\delta_{n+m,0}\cr
[A_2^n,A_3^m]=&A_{23}^{n+m}+nk_{23}\delta_{n+m,0}.\cr}
\eqn\cuarenta
$$
In \cuarenta\ $A_{13}$ and $A_{23}$ are operators, whereas
$k_{13}$ and $k_{23}$ are constants. Notice that the
brackets among the $gl(N,N)$ currents are of the form
displayed in eq. \cuarenta. Using \tseis\ together with
the derivation property (eq.\siete), it is easy to arrive
at the result
$$
\eqalign{&[ (:A_1A_2:)^n, A_3^m]=
(-1)^{g(A_2)g(A_3)}((:A_{13}A_2:)^{n+m}-mk_{13}A_2^{n+m})+
\cr &+(:A_{1}A_{23}:)^{n+m}-mA_1^{n+m}k_{23}-
\sum_{j=1-\Delta_{23}}^{m-\Delta_2}[A_1^{m+n-j},A_{23}^j],\cr}
\eqn\cuno
$$
where $\Delta_{23}$ is the dimension of the operator $A_{23}$.
The last term in the right-hand side of \cuno\
originates when one
tries to express the result in terms of
normal-ordered products.
Using  equations \cuno\ and \treinta,
we obtain after an easy
calculation  $$
[T_1^n,J_{AB}^m]=-2mkJ_{AB}^{n+m}+
2m(-1)^{g(B)}\delta_{AB}\sum_C J_{CC}^{n+m},
\eqn\cdos
$$
where we have used the fact that
$$
\sum_{A,B,C,D}(-1)^{g(B)}F_{AB,CD}^{PQ}F_{BA,RS}^{CD}=
-2(-1)^{g(R)}\delta_{PQ}\delta_{RS},
\eqn\ctres
$$
which can be easily obtained from the explicit form of the
structure constants (see eq. \doce). Eq. \ctres\ is nothing
but the value of the quadratic Casimir $C_1$ for the
adjoint representation. On the other hand eq. \cdos\
implies that $T_1$ does not act diagonally on the
$gl(N,N)$ currents. Therefore it seems plausible that we
should also consider the operator $T_2$. Again making use
of \cuno, we obtain
$$
[T_2^n,J_{AB}^m]=
-2mk(-1)^{g(B)}\delta_{AB}\sum_C J_{CC}^{n+m}.
\eqn\ccuatro
$$
{}From a glance at the right-hand sides of eqs. \cdos\ and
\ccuatro\ it is clear that we can combine $T_1$ and $T_2$
in such a way that the non-diagonal terms disappear and
$J_{AB}$ becomes a dimension-one primary field. Let us
define
$$
T={1\over 2k}T_1+{1\over 2k^2}T_2.
\eqn\ccinco
$$
{}From eqs. \cdos\ and \ccuatro\ we get
$$
[T^n,J_{AB}^m]=-mJ_{AB}^{n+m},
\eqn\cseis
$$
\ie, $J_{AB}$ is indeed primary with respect to $T$.

The form of the energy-momentum tensor $T$ displayed in eq.
\ccinco\ generalizes the result of Rozanski and Saleur [\RS],
who have obtained a similar equation for $N=1$. As
was pointed out in ref. [\RS], the
second term in \ccinco\ can
be regarded as a quantum correction to the Sugawara tensor
$T_1$. Notice that in this case we have obtained a ${1\over
k^2}$ contribution, instead of
the usual $k\rightarrow k+c_v$
shift that appears, for example, in the $SU(N)$
Wess-Zumino-Witten model\ ---$c_v$
being the quadratic Casimir
in the adjoint representation. In our case
the fact that $k$
is not shifted in $T$  originates in the vanishing of the
diagonal terms in $(C_2)^{Adjoint}$ (see eq. \ctres ), which
is due to an exact cancellation between the bosonic and
fermionic contributions in eq. \ctres.

We should also check that $T^n$ satisfies the Virasoro
algebra. With this purpose in mind, let us compute the
bracket of $T^n$ with a general quadratic
bilinear built up
with  primary fields. Suppose that $A(z)$ and $B(z)$ are
primary fields with conformal dimensions $\Delta_A$ and
$\Delta_B$ respectively. Consider the bilinear operator
$$
O(z)=:A(z)B(z):.
\eqn\csiete
$$
A calculation similar to the one that led to eq. \cuno\
gives
$$
[T^n,O^m]=[(\Delta_A+\Delta_B-1)n-m]O^{n+m}+
\sum_{q=1-\Delta_B}^{n-\Delta_B}(q-n\Delta_B)[A^{n+m-q},B^q].
\eqn\cocho
$$
Notice that $O(z)$ is a primary  operator with conformal
dimension $\Delta_A+\Delta_B$ only if the last term in the
right-hand side of \cocho\
vanishes. This is the case of $T_1$
and $T_2$,
$$
[T^n,T_1^m]=(n-m)T_1^{n+m}
\,\,\,\,\,\,\,\,\,
[T^n,T_2^m]=(n-m)T_2^{n+m},
\eqn\cnueve
$$
which, in particular, implies that
$$
[T^n,T^m]=(n-m)T^{n+m}.
\eqn\cincuenta
$$
Therefore the energy-momentum tensor operator satisfies
the Virasoro algebra without central extension.

Moreover, it follows from eqs. \vtres\ and \vseis\
that a field transforming in the fundamental
representation of the current algebra has a conformal weight
given by
$$
\Delta_{Fundamental}={1\over 2k^2}.
\eqn\ciuno
$$
Notice that only $T_2$ contributes to \ciuno. On the other
hand, due to the non-simplicity of the $gl(N,N)$ superalgebra,
a field transforming as an arbitrary representation of the
algebra is not in general a primary field of the Virasoro
algebra. This happens because, for this superalgebra, the
quadratic Casimirs evaluated in general representations are
non-diagonal (the adjoint representation provides an example
where this phenomenon occurs, see eq. \ctres).

It is interesting to stress the reason why the
Virasoro central charge $c$ of our theory vanishes. When one
applies eq. \cocho\ to obtain \cnueve, it turns out that
there is an exact compensation between the
conformal anomaly
coming from the bosonic currents and that
originated from the
fermionic ones. Actually, the conformal
field theory we are
dealing with is non-unitary, since it has an indefinite
metric in its Fock space (see, for example, the signs
appearing in the right-hand side of \tocho). Moreover, in a
$gl(N,N)$ current algebra there is an exact
balance between
fermionic and bosonic degrees of freedom
and, as a consequence,
the central charge of the Virasoro algebra vanishes. In fact,
from the results obtained in ref.
\REF\schnitzer{M. Bourdeau, E.
Mlawer, H. Riggs and  H.J. Schnitzer\journal\np&B372(92)303.}
[\schnitzer] it follows that, for
$N>M$, the $gl(N,M)$ current superalgebra shares many
properties with the $gl(N-M)$ affine (bosonic) Lie algebra. It
is thus clear that, when $N=M$, we are dealing with a limiting
case which requires a separate study. From these
considerations, one would expect the theory whose
energy-momentum tensor is given by \ccinco\ to be a
topological field theory. This is indeed
the case as we shall
show in the next section, where the nature of the theory we
have constructed will be explored in detail.

\chapter{The topological algebra}

In this section we aim at establishing the topological character
of the conformal field theory possessing the $gl(N,N)$ current
symmetry  described in the previous section. We would
like to uncover the topological symmetry of our theory and
study its relation with the original $gl(N,N)$ supersymmetry.
By a topological field theory we mean a theory in which there
exists a nilpotent (\ie\ fermionic), BRST-type symmetry
such that the energy-momentum tensor of the theory is
BRST-exact. The theory we have at hand enjoys an affine
$gl(N,N)$ symmetry and, in particular, there exist fermionic
nilpotent currents from which a topological BRST symmetry
fulfilling our requirements can be constructed. Let us see
that this is indeed the case. First of all we shall distinguish
from now on between bosonic and fermionic currents. We
introduce the following notations:
$$
\eqalign{
\Psi_{ab}\equiv &J_{a+N,b}
\,\,\,\,\,\,\,\,\,\,\,\,\,\,\,\,\,
\Lambda_{ab}\equiv J_{a,b+N}\cr
K_{ab}\equiv &J_{ab}
\,\,\,\,\,\,\,\,\,\,\,\,\,\,\,\,\,\,\,\,\,
L_{ab}\equiv J_{a+N,b+N}.\cr}
\eqn\cidos
$$
Notice that $\Psi_{ab}$ and $\Lambda_{ab}$ are fermionic (\ie\
Grassmann odd) currents, whereas $K_{ab}$ and $L_{ab}$ are
bosonic fields. With these notations the general commutation
relations (eq. \treinta) in the [boson,boson] sector take the
form
$$
\eqalign{
[K_{ab}^n,K_{cd}^m]=&
\delta_{bc}K_{ad}^{n+m}-\delta_{ad}K_{cb}^{n+m}
+kn\delta_{bc}\delta_{ad}\delta_{n+m,0}\cr
[L_{ab}^n,L_{cd}^m]=&
\delta_{bc}L_{ad}^{n+m}-\delta_{ad}L_{cb}^{n+m}
-kn\delta_{bc}\delta_{ad}\delta_{n+m,0}\cr
[K_{ab}^n,L_{cd}^m]=&0.\cr}
\eqn\citres
$$
For the [fermion, fermion] case we have
$$
\eqalign{
[\Psi_{ab}^n,\Psi_{cd}^m]=&[\Lambda_{ab}^n,\Lambda_{cd}^m]=0\cr
[\Psi_{ab}^n,\Lambda_{cd}^m]=&
\delta_{ad}K_{cb}^{n+m}+\delta_{bc}L_{ad}^{n+m}
-kn\delta_{bc}\delta_{ad}\delta_{n+m,0},\cr}
\eqn\cicuatro
$$
and finally the brackets involving both bosonic and fermionic
currents are
$$
\eqalign{
[\Psi_{ab}^n,K_{cd}^m]=&\delta_{bc}\Psi_{ad}^{n+m}
\,\,\,\,\,\,\,\,\,\,\,\,\,\,\,\,\,
[\Psi_{ab}^n,L_{cd}^m]=-\delta_{ad}\Psi_{cb}^{n+m}\cr
[\Lambda_{ab}^n,K_{cd}^m]=&-\delta_{ad}\Lambda_{cb}^{n+m}
\,\,\,\,\,\,\,\,\,\,\,\,\,
[\Lambda_{ab}^n,L_{cd}^m]=\delta_{bc}\Lambda_{ad}^{n+m}.
\cr}
\eqn\cicinco
$$
It is also interesting to write down the energy-momentum
tensor $T$ in terms of our component currents \cidos. The
contribution $T_1$ coming from the first quadratic Casimir
takes the form
$$
T_1=\sum_{a,b}:(K_{ab}K_{ba}-L_{ab}L_{ba}
+\Psi_{ab}\Lambda_{ba}-\Lambda_{ab}\Psi_{ba}):.
\eqn\ciseis
$$
Defining the current of the identity
$J_E$ as
$$
J_E=\sum_a(K_{aa}+L_{aa}),
\eqn\cisiete
$$
$T_2$ is simply given by
$$
T_2= :J_E J_E:.
\eqn\ciocho
$$
It is important to point out that from eq. \cicuatro\ it
follows that the two fermionic currents
$\Psi_{ab}$ and $\Lambda_{ab}$ present in our algebra are
nilpotent. We would like to define a BRST current without any
free index of the current algebra. The simplest way of achieving
this objective is to sum over the $gl(N)$ diagonal
components of one of the fermionic currents. Suppose we define:
$$
Q=\sum_a \Lambda_{aa}.
\eqn\cinueve
$$
Of course, nothing essential would change if we had chosen the
$\Psi$ currents instead of the $\Lambda$'s in  the definition
\cinueve. Obviously we have
$$
[Q^n,Q^m]=0.
\eqn\setenta
$$
Let us see that the current $Q$ defined above serves  our
purposes. Using eqs. \cicuatro\ and \cicinco\ we easily
obtain the brackets of $Q$ with the currents:
$$
\eqalign{
[Q^n,K_{ab}^m]=&-\Lambda_{ab}^{n+m}
\,\,\,\,\,\,\,\,\,\,\,\,\,\,
[Q^n,L_{ab}^m]=\Lambda_{ab}^{n+m}
\,\,\,\,\,\,\,\,\,\,\,\,\,\,
[Q^n,\Lambda_{ab}^n]=0\cr
[Q^n,\Psi_{ab}^m]=&L_{ab}^{n+m}+K_{ab}^{n+m}+
kn\delta_{ab}\delta_{n+m,0}.\cr}
\eqn\suno
$$
We can use the zero-mode of $Q$ to define a BRST cohomology
operator $\delta$. For any field $\Phi$ we define
$\delta\Phi$ as
$$
\delta \Phi= [Q^0,\Phi].
\eqn\sdos
$$
Taking the brackets \suno\ into account, the BRST variations
of the currents are
$$
\eqalign{
\delta K_{ab}=&-\Lambda_{ab}\cr
\delta L_{ab}=&\Lambda_{ab}\cr
\delta \Lambda_{ab}=&0\cr
\delta \Psi_{ab}=&K_{ab}+L_{ab}.\cr}
\eqn\stres
$$
Notice that the combination $K_{ab}+L_{ab}$
of the bosonic currents is $\delta$-exact. On the other hand
by a direct calculation using \stres\ and the graded
derivation property of the bracket (eq.\siete) one can check
that $T_1$ and $T_2$ are $\delta$-invariant,
$$
\delta T_1=0
\,\,\,\,\,\,\,\,\,\,\,\,\,\,\,\,\,\,\,
\delta T_2=0,
\eqn\scuatro
$$
\ie\ $T_1$ and $T_2$ are closed in the BRST cohomology defined
by $\delta$. Actually, as we shall prove below, $T_1$ and $T_2$
(and thus the complete energy-momentum tensor) are
cohomologically exact. In order to verify this fact it is
clear that we must find fermionic operators whose BRST
variations give $T_1$ and $T_2$. We would like to have an
expression of these odd partners of $T_1$ and $T_2$ in terms
of the underlying currents of the model. The natural way to
proceed is to imitate the Sugawara construction employed to
obtain the energy-momentum tensor and consider operators
that are bilinear in the currents. Due to the fermionic
character of the operators we are looking for, it is clear
that we must deal with  products of a
fermionic and a bosonic current evaluated at the same
point. After some trial and error, we easily arrive at the
desired result. In fact if we define
$$
G_1=\sum_{a,b}:(\Psi_{ab}K_{ba}-L_{ab}\Psi_{ba}):,
\eqn\scinco
$$
it is easy to check that
$$
\delta G_1=T_1.
\eqn\sseis
$$
In the same way, inspecting eqs. \stres, \cisiete\ and
\ciocho, we conclude
$$
T_2=\delta G_2,
\eqn\ssiete
$$
with
$$
G_2=\sum_a :\Psi_{aa}J_E:.
\eqn\socho
$$
Therefore if we define $G$ as
$$
G={1\over 2k}G_1+{1\over 2k^2}G_2,
\eqn\snueve
$$
we have
$$
T=\delta G,
\eqn\setenta
$$
\ie\ $G$ is the odd partner of $T$  we were looking for. Let
us now verify that this new fermionic field $G$ is primary
 with respect to  $T$ with conformal weight equal to two.
This fact can be established by computing the bracket
$[T^n,G^m]$. Using our general expression \cocho\ for general
bilinear operators, we immediately get
$$
[T^n,G^m]=(n-m)G_{n+m},
\eqn\seuno
$$
which proves the statement.

Once we have obtained the explicit form of $G$, we may ask
ourselves what is the algebra that $G$ closes with $T$ and
$Q$. We shall refer to it as  the topological algebra. It
is interesting to point out that eq. \setenta\ only
determines the bracket of the zero-mode of $Q$ with the
Laurent modes of $T$. As we want to realize our topological
BRST symmetry as a local symmetry, we should worry about the
general bracket $[Q^n,G^m]$. On general grounds we can write
$$
[Q^n,G^m]=T^{n+m}+nR^{n+m}+{d\over 2}m(m+1)\delta_{n+m,0},
\eqn\sedos
$$
where $d$ is a c-number and $R$ is a dimension-one abelian
current. Actually eq. \sedos\ is the bracket
obtained when the $N=2$ superconformal algebra is twisted
[\EY,\dij]. This twist consists in a redefinition of the
energy-momentum tensor of the superconformal theory by adding
the derivative of the $U(1)$ current $R$ appearing in the
$N=2$ superconformal algebra:
$$
T=T_{N=2}+{1\over 2}\partial R.
\eqn\setres
$$
After this redefinition eq. \sedos\  is satisfied\ ---$Q$ and
$G$ being the two supersymmetry currents of the initial
theory. The Virasoro central charge for the twisted model
vanishes and thus the redefined energy-momentum tensor
satisfies \cincuenta. Moreover, the c-number anomaly
$d$ appearing in the right-hand side of \sedos\ is related to
the central charge of the untwisted theory as
$$
d={ C_{N=2}\over 3}.
\eqn\secuatro
$$
Other brackets obtained by twisting the $N=2$
superconformal algebra are
$$
\eqalign{
[T^n,R^m]=&-mR^{n+m}-{d\over 2}n(n+1)\delta_{n+m,0}\cr
[T^n,Q^m]=&-mQ^{n+m}\cr
[R^n,R^m]=&dn\delta_{n+m,0}\cr
[R^n,G^m]=&-G^{n+m}\cr
[R^n,Q^m]=&Q^{n+m}.\cr}
\eqn\secinco
$$
Furthermore eq. \seuno\ also holds. Actually, in the original
superconformal theory, both $Q$ and $G$ behave as primary
fields of conformal dimension ${3\over 2}$. After the
redefinition \setres, $Q$ becomes a dimension-one operator,
whereas $G$ acquires conformal weight $2$. This different
behaviour is due to the different $R$-charges of these two
operators (see eq. \secinco).

The theory we arrive at by twisting is a model of topological
matter. In case we start with a conformal invariant
sigma model, $d$ is
nothing but the dimension of the target space in which the
bosonic sector of the theory is embedded. For this reason we
shall call $d$ from now on the dimension of the topological
algebra. Let us also recall [\witgrav] that, when topological
matter is coupled to topological gravity, the resulting theory
reproduces many features of the matrix models that describe
non-critical strings.

In our $gl(N,N)$ theory eqs. \sedos\ and \secinco\ are
satisfied for $d=0$, \ie
$$
d_{gl(N,N)}=0.
\eqn\seseis
$$
In order to check \seseis\ one has to use the explicit
expressions of $Q$, $G$ and $T$, together with the basic
brackets (eqs. \citres\ -- \cicinco) in our general equation
for the brackets of an arbitrary bilinear operator (eq.
\cuno). In so doing, one also obtains the explicit form of
the $R$ current, which turns out to be
$$
R={1\over 2}\sum_a(K_{aa}-L_{aa}).
\eqn\sesiete
$$
It is important to emphasize that, in our approach, we do not
have to perform a modification of the supersymmetric theory
we started with in order to obtain a topological field
theory, \ie\ no twist is needed. Actually, as we shall now
show, the topological algebra of our $gl(N,N)$ theory differs
from the one obtained from the $N=2$ superconformal models.
In fact the topological algebra of the former can be regarded
as an extension of  the latter. The difference between
these two algebras shows up when computing the bracket
$[G^n,G^m]$, which vanishes in the twisted $N=2$ algebra. The
calculation of this bracket in our case is rather involved,
so let us first give some of  the intermediate steps, which
will turn out to be very useful in what follows; in fact these
results are interesting by themselves. First of all we compute
the brackets among the odd companion $G$ of the energy-momentum
tensor and the currents of the model, which can be obtained as
particular cases of our general equation \cuno. Since $G$ is
the sum of two contributions, we have to consider
brackets involving these two terms separately. First of all we
write down
the brackets of $G_1$ and the bosonic  currents,
$$
\eqalign{
[G_1^n,K_{ab}^m]=&{\cal G}_{1,ab}^{n+m}+m\delta_{ab}\sum_c
\Psi_{cc}^{n+m}-mk\Psi_{ab}^{n+m}\cr
[G_1^n,L_{ab}^m]=&-{\cal G}_{1,ab}^{n+m}-m\delta_{ab}\sum_c
\Psi_{cc}^{n+m}-mk\Psi_{ab}^{n+m},\cr}
\eqn\seocho
$$
where ${\cal G}_{1,ab}$ is a fermionic bilinear operator
given by
$$
{\cal G}_{1,ab}=:\sum_c (\Psi_{ac}K_{cb}-L_{ac}\Psi_{cb}):.
\eqn\senueve
$$
In the same way, the brackets of $G_1$ and the fermionic
currents $\Psi$ and $\Lambda$ are given by
$$
\eqalign{
[G_1^n,\Psi_{ab}^m]=&0\cr
[G_1^n,\Lambda_{ab}^m]=&{\cal T}_{1,ab}^{n+m}+
km(K_{ab}^{n+m}-L_{ab}^{n+m})
-m\delta_{ab}J_E^{n+m},\cr}
\eqn\ochenta
$$
with
$$
{\cal T}_{1,ab}=\sum_c:(K_{ac}K_{cb}-L_{ac}L_{cb}
+\Psi_{ac}\Lambda_{cb}-\Lambda_{ac}\Psi_{cb}):.
\eqn\ouno
$$
The brackets involving $G_2$ are calculated in the same way,
with the result
$$
\eqalign{
[G_2^n,K_{ab}^m]=&{\cal G}_{2,ab}^{n+m}-mk\delta_{ab}\sum_c
\Psi_{cc}^{n+m}\cr
[G_2^n,L_{ab}^m]=&-{\cal G}_{2,ab}^{n+m}+mk\delta_{ab}\sum_c
\Psi_{cc}^{n+m}\cr
[G_2^n,\Psi_{ab}^m]=&0\cr
[G_2^n,\Lambda_{ab}^m]=&{\cal T}_{2,ab}^{n+m}+
mk\delta_{ab}J_E^{n+m},\cr}
\eqn\odos
$$
where now ${\cal G}_{2,ab}$ and ${\cal T}_{2,ab}$ are given
by
$$
\eqalign{
{\cal G}_{2,ab}=&:\Psi_{ab}J_E:\cr
{\cal T}_{2,ab}=&:(K_{ab}+L_{ab})J_E:.\cr}
\eqn\otres
$$
Notice that $G_i=\sum_{a,b}{\cal G}_{i,ab}$ and
$T_i=\sum_{a,b}{\cal T}_{i,ab}$, for $i=1,2$.

Using the basic results \seocho, \ochenta\ and \odos\ we can
obtain the brackets between the different terms of $G$. After
some calculations we get
$$
\eqalign{
[G_1^n,G_1^m]=&2\sum_{a,b}(:{\cal G}_{1,ab}\Psi_{ba}:+
:\Psi_{aa}\partial \Psi_{bb}:)^{n+m}\cr
[G_1^n,G_2^m]=&[G_2^n,G_1^m]=2k\sum_{a,b}
(:\partial \Psi_{aa}\Psi_{bb}:)^{n+m}\cr
[G_2^n,G_2^m]=&0.\cr}
\eqn\ocuatro
$$
In order to get \ocuatro\ one has to pursue the same steps we
followed to obtain eq. \cuno. Notice that this last equation
is not applicable here, since the brackets \seocho, \ochenta\
and \odos\ are not of the form displayed in eq. \cuarenta.
Taking  eq. \senueve\ into account, we arrive at
$$
[G^n,G^m]=W^{n+m},
\eqn\ocinco
$$
where $W$ is a bosonic operator, trilinear in the currents,
whose explicit expression is
$$
W={1\over 2k^2}:Tr({\cal G}_1\Psi):
+{1\over 2k^2}:Tr\partial \Psi Tr\Psi:.
\eqn\oseis
$$
In eq. \oseis\ we have used a trace notation to represent the
double $gl(N,N)$ summation of eq. \ocuatro; this notation
simplifies  our equations greatly and will be frequently used
from now on.

As  was announced above, $W$ does not vanish in general.
Notice that $W$ is a spin-three operator which, as we shall
show below,  behaves as a Virasoro primary field. The
presence of this $W$ operator in the right-hand side of
\ocinco\ will force us to extend the topological algebra. In
this extended algebra one should include the brackets of $W$
with all other generators (\ie\ with $Q$, $R$, $T$
and $G$). In principle this process will introduce new
fields and there is no guarantee that the algebra will close
with a finite number of generators. A priori, this situation is
similar to the one presented when one analyses
$W$-algebras which, in general, only close with a
finite number of fields if they are non-linear. However, in our
case, the situation is quite different. We shall check below
that, in order to close the algebra, we will  need to
introduce just one  additional operator (a BRST partner of
$W$) and nevertheless the algebra will remain linear. Before
launching into the calculations supporting this conclusion,
let us study in what cases $W$ is not identically zero.
Having this purpose in mind, we shall reorder $W$
using the rules developed in the previous section. The
basic idea consists in getting the fermionic currents
appearing in the trilinear expression of $W$ to contiguous
places; in so doing, we shall be able to use in some cases the
nilpotency of these currents to conclude the vanishing of the
corresponding contribution to $W$. Using \tsietei\ we can rewrite
the first term of eq. \oseis\ as\ :
$$
:Tr({\cal G}_1\Psi):=
\sum_{a,b,c}:(\Psi_{ac}K_{cb}\Psi_{ba}-
L_{ac}\Psi_{cb}\Psi_{ba}):.
\eqn\osiete
$$
Let us now reorder the first term in the right-hand side of
\osiete\  using the general equation \tsietei. Taking  the
bracket of $K$ and $\Psi$ displayed in eq. \cicinco\ into
account, we can readily write:
$$
\sum_{a,b,c}:\Psi_{ac}K_{cb}\Psi_{ba}:=-
\sum_{a,b,c}:K_{ab}\Psi_{bc}\Psi_{ca}:-
N\sum_{a,b}:\partial \Psi_{ab}\Psi_{ba}:.
\eqn\oocho
$$
This result can be plugged back in eq. \osiete, after which
we get
$$
W=-{1\over 2k^2}:Tr[(K+L)\Psi\Psi]:+
{1\over 2k^2}:[Tr\partial\Psi Tr\Psi-N Tr(\partial \Psi\Psi)]:.
\eqn\onueve
$$
This reordered expression shows that  $W$ vanishes in the
$gl(1,1)$ theory. Indeed, in this case we only have a single
$\Psi$ current and therefore we can eliminate the trace
operation appearing in \onueve. For $N=1$, due to the nilpotency
of $\Psi$, the first term in the right-hand side of \onueve\
vanishes, while the other two terms cancel with each other.
Thus, only for $gl(N,N)$ with $N>1$ do we really have a
topological algebra extended by a dimension-three operator. Let
us now come back to the general case. An important piece of
information is obtained by computing the BRST variation of $W$.
Concentrating on the first term in \oseis, which is
proportional to  $:Tr({\cal G}_1\Psi):$ and using the BRST
variation of the currents (eq. \stres), we get
$$
\delta Tr (:{\cal G}_1\Psi:)=
:Tr [{\cal T}_1\Psi-{\cal G}_1(K+L)]:.
\eqn\noventa
$$
Freely reordering  the fields in the right-hand side of eq.
\noventa\ we get a complete cancellation of all the terms
trilinear in the currents. However, we know that with this
reordering we generate (as in eq. \tsietei) bilinear terms
containing the derivatives of the fields. In our case we are
left with
$$
\eqalign{
\delta Tr (:{\cal G}_1\Psi:)=&-:Tr(\partial K+\partial L)
Tr \Psi :+:Tr\partial \Psi Tr(K+L): \,=\cr
=&-\delta (Tr:\partial \Psi \Psi:),\cr}
\eqn\nuno
$$
where in the last step we have used  eq. \stres\ again to
express the result as a total $\delta$-variation. Comparing
eq. \nuno\ with the expression of $W$ (eq. \onueve), we
conclude that $W$ is $\delta$-closed, namely,
$$
\delta W=0.
\eqn\ndos
$$
Remarkably $W$ is also $\delta$-exact. It is easy to guess
the expression whose BRST variation gives $W$. For this
purpose eq. \onueve\ is very illustrative. Consider the first
term in this expression. Given that
$\delta \Psi=K+L$, one is led to suppose that $\delta\Psi^3$
is proportional to $W$ and that the last two terms in
\onueve\  containing derivatives of $\Psi$ will originate as
the result of the change of order of the fields inside the
normal-ordered product. That this is indeed the case can be
checked by a direct calculation that we now describe. Using
\stres\ we get
$$
\delta Tr(:\Psi^3:)=
:Tr[(K+L)\Psi\Psi-\Psi(K+L)\Psi+\Psi\Psi(K+L)]:.
\eqn\ntres
$$
The three terms in eq. \ntres\ have the same structure. If we
try to convert the second and third term into the first one
(which has the order appearing in the expression of $W$) we
generate derivative terms terms as follows:
$$
{1\over 3}\delta Tr(:\Psi^3:)=:Tr[(K+L)\Psi\Psi]:-
:Tr\partial \Psi Tr\Psi:+N:Tr(\partial \Psi\Psi):.
\eqn\ncuatro
$$
Comparing eqs. \ncuatro\ and \onueve, it is evident that
upon defining the fermionic operator $V$ as
$$
V\equiv -{1\over 6k^2}Tr(:\Psi^3:),
\eqn\ncinco
$$
one has
$$
W=\delta V,
\eqn\nseis
$$
which is the desired result. Notice that for $gl(1,1)$
current algebras $V$ vanishes identically, which is
consistent with the fact that $W$ is also zero in this case.

One might wonder if there is a redefinition of $T$, $G$, and $R$
such that the algebra they close is just the twisted $N=2$
superconformal algebra, \ie, such that $G$ is nilpotent and $W$
can be set to zero. In other words, we are asking ourselves to
what extent is it unavoidable to deal with the extended algebra
instead of the usual (non-extended) one. We could try, for
example, to exploit the fact that the choice for the BRST partner
of $T$ is not unique. Indeed, we could add to $G$ a BRST
variation of a dimension-two bosonic operator. It is easy to
convince oneself that, if we want to keep the expressions for
our generators local in the currents, then there is only one
possibility for $G$ (\ie\ the one we have chosen in eq.
\snueve). As our $Q$-symmetry carries $R$-charge $+1$  and $G$
must have $R$-charge $-1$ in the topological algebra (see eq.
\secinco), we need a dimension-two bosonic operator with
$R$-charge equal to $-2$ in order to preserve the $R$-symmetry.
Moreover, we can only add  $sl(N)$ singlet operators to $G$.
Taken the fact that the bosonic currents $K$ and $L$ have
vanishing $R$-charge, and that the $\Lambda$'s possess
$R$-charge equal to $+1$, we are forced to consider operators
constructed with two $\Psi$ currents. So the only possible
terms satisfying all the above conditions and whose
$\delta$-variations can be added to $G$ are of the form
$Tr(\Psi^2 )$ and $Tr(\Psi )Tr(\Psi )$, which vanish identically
due to eq. \cicuatro.

By inspecting eqs. \setenta\ and \nseis\ one immediately
notices a clear parallelism. Inspired by this analogy one
could be led to think that the implementation of eq.
\nseis\ by means of the local current $Q(z)$ would oblige us to
consider new operators, which would show up when $V$ is
anticommuted with the non-zero modes of $Q$. Fortunately this
is not the case, as we are now going to establish, and our
extended topological algebra closes with the only addition of
$W$ and its BRST partner $V$ to the generators $T$, $G$, $R$,
and $Q$ of the twisted $N=2$ superconformal algebra.

As we have an explicit representation of the operators in
terms of the $gl(N,N)$ currents, we can compute  the
brackets involving $V$ and $W$ directly. However in some cases
this direct calculation is rather long and tedious. Fortunately
the same result can be obtained in a much simpler way by
exploiting consistency conditions of the brackets already
obtained explicitly. The fact that in the bracket of $G$
with itself only  one operator (\ie\ $W$) appears is crucial
in order to solve these consistency conditions. The
generalized Jacobi identity \seis\ will become the basic tool
in this approach. For example, if we calculate the bracket of
$Q^r$ with both sides of eq.\ocinco\  and apply \seis, we
obtain
$$
[Q^r,W^{n+m}]=-[G^n,[G^m,Q^r]]-[G^m,[Q^r,G^n]].
\eqn\nsiete
$$
After making use of eqs. \sedos, \seuno\ and \secinco\ in the
right-hand side of \nsiete\ we arrive at
$$
[Q^r,W^n]=0,
\eqn\nocho
$$
which generalizes eq. \ndos.

In the same way we can prove that $W$ is a Virasoro primary
field with conformal dimension three. We start by computing
the bracket of $T^r$ with both sides of eq. \ocinco. Using
again the Jacobi identity and the Virasoro primary character
of $G$ (eq. \seuno) we get
$$
\eqalign{
[T^r,W^{n+m}]=&-[G^n,[G^m,T^r]]+[G^m,[T^r,G^n]]=\cr
=&(r-m)[G^n,G^{m+r}]+(r-n)[G^m,G^{n+r}],\cr}
\eqn\nnueve
$$
which, after using \ocinco\ again, yields the desired result:
$$
[T^r,W^n]=(2r-n)W^{r+n}.
\eqn\cien
$$

Let us now study the behaviour of $W$ under the $R$-symmetry.
Computing the bracket of $R$ with eq. \ocinco\ and proceeding
as above we obtain:
$$
[R^r,W^{n+m}]=-[G^n,G^{m+r}]-[G^m,G^{n+r}].
\eqn\cieuno
$$
If we now take  the basic eq. \ocinco\ into account, we see
that $W$ has charge $-2$ with respect to the abelian current
$R$:
$$
[R^r,W^n]=-2W^{r+n},
\eqn\ciedos
$$
On the other hand from \seis\ we have
$$
[Q^r,[G^n,W^m]]=[G^n,[W^m,Q^r]]-[W^m,[Q^r,G^n]],
\eqn\cietres
$$
and using eqs. \nocho, \sedos, \cien\ and \ciedos\ we get:
$$
[Q^r,[G^n,W^m]]=(2n-m)W^{n+m+r}.
\eqn\ciecuatro
$$
Putting $r=0$ in \ciecuatro\ we obtain the BRST variation of
$[G^n,W^m]$
$$
\delta ([G^n,W^m])=(2n-m)W^{n+m}.
\eqn\ciecinco
$$
Comparing this last equation with \nseis\ one is tempted to
write
$$
[G^n,W^m]=(2n-m)V^{n+m}.
\eqn\cieseis
$$
This equation actually holds as we shall prove  below
using an independent argument. Let us assume for a moment
that eq. \cieseis\ is satisfied and see what  the
consequences are. Substituting back this result into \ciecuatro\
we get:
$$
[Q^r,V^n]=W^{r+n}.
\eqn\ciesiete
$$
It is instructive to compare this last equation with eq.
\sedos\ relating the energy-momentum tensor to its BRST
partner. As was stated previously, contrary to what
happens with $T$ and $G$, the relation between $W$ and its
fermionic counterpart $V$ does not involve any new operator.

There are still several other brackets which we have to
compute in order to completely check  the closure of the
topological algebra. First of all, some of them
can be shown to vanish by a simple inspection of their explicit
form. For example, taking into account that $\Psi$
anticommutes with itself and that
$$
[{\cal G}_{1,ab}^n,\Psi_{cd}^m]=0,
\eqn\cieocho
$$
we can write
$$
\eqalign{
[V^r,V^n]=&0\cr
[V^r,W^n]=&0\cr
[G^r,V^n]=&0,\cr}
\eqn\cienueve
$$
where in order to prove the last equality we must use eqs.
\ochenta\ and \odos. Moreover it is
a straightforward consequence of our general equation \cocho\
that
$$
[T^r,V^n]=(2r-n)V^{r+n},
\eqn\ciediez
$$
\ie , $V$ is a dimension-three primary field.

Let us now  give the promised proof of eq. \cieseis.
First of all, using eqs. \setenta\ and \nseis\ we have
$$
\delta ([G^r,V^n])=[T^r,V^n]-[G^r,W^n].
\eqn\cieonce
$$

By virtue of the last equation in \cienueve, the left-hand side
of \cieonce\ vanishes. The primary character of $V$ with
respect to $T$ (eq. \ciediez ) is easily seen to imply
\cieseis. Now,  using eqs. \ndos\ and \nseis\ it
follows that
$$
\delta([V^r,W^n])=[W^r,W^n].
\eqn\ciedoce
$$
And, after using the second equation in \cienueve, we conclude
that
$$
[W^r,W^n]=0.
\eqn\cietrece
$$

In order to complete the algebra it remains to compute the
bracket of $R$ with $V$. The Jacobi identity implies:
$$
[R^r,[G^n,W^m]]=-[G^n,[W^m,R^r]]-[W^m,[R^r,G^n]].
\eqn\ciecatorce
$$
As $G$ and $W$ have $R$-charges $-1$ and $-2$ (see eqs.
\secinco\ and \ciedos) it follows that
$$
[R^r,[G^n,W^m]]=-2[G^n,W^{m+r}]+[W^m,G^{n+r}].
\eqn\ciequince
$$
Using \cieseis\ in both sides of \ciequince\ one readily
concludes that
$$
[R^r,V^n]=-3V^{r+n},
\eqn\ciedseis
$$
\ie, $V$ has $R$-charge equal to $-3$. This bracket
completes the set of relations defining the extended
topological algebra.  We have gathered all the
brackets we have obtained in Appendix A, where we have written the
form of the algebra for arbitrary dimension $d$. Up to now, we
have obtained a representation for $d=0$ only. In the next
section we shall see that there exists a modification of the
generators giving rise to a $d\ne 0$ extended algebra.

As  was mentioned in section 1, the extended topological
algebra we have found has been previously studied by
Kazama [\kaza], who characterized algebraically some
non-trivial extensions of the twisted $N=2$ superconformal
algebra. The algebra compiled in Appendix A was obtained
form a truncation of a larger algebra in which there are two
extra operators, one fermionic and the other bosonic, of
dimensions zero and one respectively. This larger extended
algebra is satisfied by the $bc$ system studied by
Distler [\dist], which is relevant in two-dimensional quantum
gravity. Nevertheless, as far as  we know, an explicit
representation of the algebra displayed in  Appendix A was not
known.

To summarize, in this section we have been able to obtain the
 algebra  encoding the topological symmetry of
our $gl(N,N)$ theory. The generators of this algebra have
dimensions 1, 2, and 3 and have been represented as
normal-ordered products of currents. Only for the $gl(1,1)$
case does our model possess a twisted $N=2$ superconformal
algebra. In general we have to deal with an extended topological
algebra, which has the important property of closing with a
set of $gl(N)$ bosonic and fermionic currents. This last
question will be further explored in the next section.

\chapter{The topological current algebra and $d \not = 0$
extensions}

Let us now discuss the relation of our topological algebra
with the underlying current algebra from which the former
has been obtained. The brackets of $T$ and the different
currents are fixed by the fact that the $J_{AB}$ are primary
dimension-one operators (see eq. \cseis). It would therefore
seem natural to expect the (anti)commutator of $G$ with the
currents to be given by an odd
analogue of eq. \cseis. However this is not the case, as  can
be verified by combining eqs. \seocho, \ochenta\ and \odos.
One has
$$
\eqalign{
[G^n,K_{ab}^m]=&{\cal G}_{ab}^{n+m}-{m\over
2}\Psi_{ab}^{n+m}\cr [G^n,L_{ab}^m]=&-{\cal
G}_{ab}^{n+m}-{m\over 2}\Psi_{ab}^{n+m}\cr
[G^n,\Psi_{ab}^m]=&0\cr [G^n,\Lambda_{ab}^m]=&{\cal
T}_{ab}^{n+m}+ {m\over 2}(K_{ab}^{n+m}-L_{ab}^{n+m}),\cr}
\eqn\ciedsiete
$$
where
$$
\eqalign{
{\cal G}_{ab}=&{1\over 2k}{\cal G}_{1,ab}
+{1\over 2k^2}{\cal G}_{2,ab}\cr
{\cal T}_{ab}=&{1\over 2k}{\cal T}_{1,ab}
+{1\over 2k^2}{\cal T}_{2,ab}.\cr}
\eqn\ciedocho
$$
Therefore we conclude that, contrary to what happens with
$T$, the algebra of its topological partner $G$ and the
currents does not close, since new dimension-two operators
(${\cal G}_{ab}$ and ${\cal T}_{ab}$ ) appear in the
right-hand side of \ciedsiete. These new fields have to be
introduced in the algebra and when this is done we get a
proliferation of additional operators. Performing a
similar calculation with $V$ and $W$ we would reach an equivalent
conclusion. For example the bracket $[W,L_{ab}]$ gives rise
to new dimension-three operators. Quite
remarkably, however,  there is a
subset of currents that closes with  the extended topological
algebra without introducing new fields. Consider first of all
the bosonic currents. By inspecting \ciedsiete\ one readily
realizes that the diagonal combination
$$
{\cal J}_{ab}\equiv K_{ab}+L_{ab}
\eqn\ciednueve
$$
satisfies
$$
[G^n,{\cal J}_{ab}^m]=-m\Psi_{ab}^{n+m},
\eqn\cieveinte
$$
which is indeed the odd analogue of
$$
[T^n,{\cal J}_{ab}^m]=-m{\cal J}_{ab}^{n+m}.
\eqn\cievuno
$$
Thus $G$ closes with  half  the bosonic currents. The
situation for the fermionic ones is much simpler. It is
evident that $G$  closes with the $\Psi$'s only. Recall (see
eq. \stres) that the $\Psi$'s are the fermionic partners of
the ${\cal J}$'s. In fact using \suno\ we can write
$$
\eqalign{
[Q^n,\Psi_{ab}^m]=&{\cal J}_{ab}^{n+m}+kn\delta_{ab}\delta_{n+m,0}\cr
[Q^n,{\cal J}_{ab}^m]=&0.\cr}
\eqn\cievdos
$$
Eq. \cievdos\ relating $\Psi$ to its BRST partner ${\cal J}$
has a close resemblance to eq. \sedos\  relating $Q$
and $T$. This similarity is reinforced when one looks at the
algebra closed by ${\cal J}$ and $\Psi$:
$$
\eqalign{
[{\cal J}_{ab}^n,{\cal J}_{cd}^m]=&\delta_{bc}{\cal J}_{ad}^{n+m}
-\delta_{ad}{\cal J}_{cb}^{n+m}\cr
[\Psi_{ab}^n,{\cal J}_{cd}^m]=&\delta_{bc}\Psi_{ad}^{n+m}
-\delta_{ad}\Psi_{cb}^{n+m}\cr
[\Psi_{ab}^n,\Psi_{cd}^m]=&0.\cr}
\eqn\cievtres
$$
Notice that no central extension  appears in \cievtres. In
fact the first bracket in \cievtres\ is the basic commutator
of an affine, zero level, $gl(N)$ current algebra. We can
summarize the situation as follows: $(G,T)$ and $(\Psi, {\cal
J}$) are topological doublets of operators with dimensions two
and one respectively. In both cases the two members of this
topological multiplet close an algebra without any central
element. However a c-number anomaly is obtained when the
current of the topological symmetry is anticommuted with the
fermionic member of the doublet (of course, in the case of the
dimension-two doublet $(G,T)$, the abelian current $R$ is also
obtained when $Q$ acts on $G$).
Based on this similarity,
if a system is endowed with  bosonic and fermionic
currents satisfying relations like eqs. \cievdos\ and
\cievtres, we will say that it possesses a topological current
algebra, in the same sense that the algebra in Appendix A is the
topological version of the Virasoro algebra.
This type of algebras are present in other
topological theories. This fact is illustrated in Appendix B,
where an $sl(2)$ WZW model at zero level  is studied as a
topological field theory
\REF\yos{H. Yoshii \journal \pl&B275(92)70.}
[\yos].

For completeness we should compute the bracket of all the
generators appearing in the topological algebra with the
currents. An easy calculation gives how ${\cal J}$ and $\Psi$
behave under the $R$ symmetry:
$$
\eqalign{
[R^n,{\cal J}_{ab}^m]=&kn\delta_{ab}\delta_{n+m,0}\cr
[R^n,\Psi_{ab}^m]=&-\Psi_{ab}^{n+m}.\cr}
\eqn\cievcuatro
$$
Finally, by simple inspection we get:
$$
[W^n,\Psi_{ab}^m]=[V^n,\Psi_{ab}^m]=0.
\eqn\cievcinco
$$
Performing a $\delta$-variation of this equation and using
eqs. \ndos, \nseis\ and \stres\ one obtains
$$
[W^n,{\cal J}_{ab}^m]=[V^n,{\cal J}_{ab}^m]=0.
\eqn\cievseis
$$
Altogether eqs. \cieveinte -- \cievseis\ imply that our
conformal topological algebra is compatible with the $gl(N)$
topological current algebra generated by ${\cal J}_{ab}$ and
${\Psi_{ab}}$.

As  was proved above, our $gl(N,N)$ theory satisfies the
extended topological algebra with the parameter $d$ equal to
zero (see eq. \seseis). Let us ask ourselves if it is possible
to modify the theory in such a way that it continues to be
topological but the parameter $d$ does not vanish. In other
words, we would like to find out whether or not there are
marginal directions (in the moduli space of topological
theories to which our model belongs) such that one can move
away from the $d=0$ point. With this objective in mind, notice
that we can deform our energy-momentum tensor by adding a
two-dimensional term that preserves its  BRST-exactness.
Indeed, due to the last equality in eq. \stres, it follows
that for any c-number constant matrix $\alpha_{ab}$, the
transformation
$$
T\rightarrow T+\sum_{a,b}\alpha _{ab}\partial {\cal J}_{ab}
\eqn\cievsiete
$$
keeps the operator $T$ $\delta$-exact. Moreover the transformed
operator $T$ satisfies the extended topological algebra if $G$
and $R$ are modified as
$$
\eqalign{
G&\rightarrow G+\sum_{a,b}\alpha _{ab}\partial \Psi_{ab}\cr
R&\rightarrow R+\sum_{a,b}\alpha _{ab}{\cal J}_{ab}\cr}
\eqn\cievocho
$$
and the spin-three operators $V$ and $W$ remain unchanged. It is
important to stress the fact that, under the transformations
\cievsiete\ and \cievocho, the algebra retains its form, the
only modification introduced being the change of the parameter
$d$ (see below). In
terms of modes, the twist performed in eqs. \cievsiete\ and
\cievocho\ is equivalent to
$$
\eqalign{
T^n&\rightarrow T^n-(n+1)\sum_{a,b}\alpha _{ab}{\cal
J}_{ab}^n\cr
G^n&\rightarrow G^n-(n+1)\sum_{a,b}\alpha _{ab} \Psi_{ab}^n\cr
R^n&\rightarrow R^n+\sum_{a,b}\alpha _{ab}{\cal
J}_{ab}^n.\cr}
\eqn\cievnueve
$$
The dimension of the modified algebra now becomes
$$
d=2kTr(\alpha).
\eqn\cietreinta
$$
Therefore, in principle we have (at least) $N^2$ possible
modular parameters in the space of topological theories
related to our original $gl(N,N)$ model. This number of
parameters is drastically reduced if we require to our twisted
theory to
possess a current algebra symmetry. Consider first the
bosonic currents. An easy calculation shows that the
commutators of the modified operator $T$ with $K$ and $L$ are
given by
$$
\eqalign{
[T^n,K_{ab}^m]=&-mK_{ab}^{n+m}+(n+1)\left( \sum_c
(\alpha_{bc}K_{ac}^{n+m}-\alpha_{ca}K_{cb}^{n+m})
-kn\alpha_{ba}\delta_{n+m,0}\right)\cr
[T^n,L_{ab}^m]=&-mL_{ab}^{n+m}+(n+1)\left( \sum_c
(\alpha_{bc}L_{ac}^{n+m}-\alpha_{ca}L_{cb}^{n+m})
+kn\alpha_{ba}\delta_{n+m,0}\right).\cr}
\eqn\cietuno
$$
In view of this result it is clear that the only chance to
have a $gl(N)$ current algebra is by requiring the second
term in the right-hand side of \cietuno\ to vanish. In general
this only takes place when the matrix $\alpha$ is proportional
to the unit matrix. Accordingly let us write
$$
\alpha_{ab}=\alpha\delta_{ab}.
\eqn\cietdos
$$
With this restriction we immediately obtain from \cietuno\
that the currents ${\cal J}$ appearing in the topological
current algebra are primary dimension-one operators for the
twisted energy-momentum operator. However the
combinations $K_{ab}-L_{ab}$ of the bosonic currents pick up
an anomaly which is proportional to the dimension $d$ of the
algebra. According to eq. \cietreinta\ this parameter $d$ is
now equal to $2k\alpha$. A straightforward calculation shows
that when eq. \cietdos\ holds, the $\Psi_{ab}$ are also primary
and eq. \cieveinte --\cievseis\ are satisfied. Therefore the
only remaining variable $\alpha$ parametrizes a line in the
moduli space of conformal topological theories, such that all
models contained in this line enjoy the $gl(N)$ topological
current algebra described above. From this viewpoint $\alpha
=0$ corresponds to a model in which this symmetry is enhanced
and we have a full  set of $gl(N,N)$ conserved currents.
Notice that, when $\alpha_{ab}$ has the form displayed in eq.
\cietdos, $T$, $R$ and $G$ are twisted along the direction of the
unit of the $gl(N)$ algebra (see eqs. \cievsiete\ and
\cievocho). In order to fix this deformation parameter
$\alpha$, one needs to impose further requirements. For
example, for the $gl(1,1)$ model, the standard minimal
topological matter
\REF\li{K. Li
\journal\np&354(91)711 \journal\np&354(91)725.}[\li] with
dimension ${n\over n+2}$,  $n$ being an arbitrary integer, is
obtained by putting $\alpha={n\over 2k(n+2)}$. In the next
section we shall check, using a free field representation, that
in this case our theory is equivalent to a twisted $N=2$
minimal superconformal model. In the general $gl(N,N)$ case a
criterion to fix $\alpha$ is lacking.

As pointed out by Kazama [\kaza], a hint to understand the
nature of our extended topological algebra is provided by the
observation that it can be obtained by twisting an $N=1$
superconformal model in which extra supermultiplets are
present. In order to check this point, let us define the
twisted energy-momentum tensor $\widetilde T$ as follows:

$$
\widetilde T=T-{1\over 2}\partial R.
\eqn\ciettres
$$
Notice that this twist is performed along a $U(1)$ direction
orthogonal to the identity current $J_E$, which corresponds to
the twist \cievsiete\ when the eq. \cietdos\ is imposed. Another
difference between \cievsiete\ and \ciettres\ is that in this
latter case there is no free parameter. It is an easy exercise
to prove that the modes of the operator $\widetilde T$ satisfy
the commutation relations
$$
[\widetilde T^n,\widetilde T^m]=(n-m)\widetilde T^{n+m}+
{d\over 4}n(n^2-1)\delta_{m+n,0}
,\eqn\cietcuatro
$$
which correspond to a central charge
$$
c=3d.
\eqn\cietcinco
$$
The coefficient in the second term of \ciettres\ has been
chosen in such a way that, with respect to the modified tensor
$\widetilde T$, the $U(1)$ current $R$ becomes a primary
dimension-one operator. Let us define the
following  operators:
$$
\eqalign{
\widetilde G^n=& G^{n-{1\over2}}
\,\,\,\,\,\,\,\,\,\,\,\,\,\,\,\,\,\,\,
\widetilde Q^n=Q^{n+{1\over2}}\cr
\widetilde V^n=& V^{n-{3\over2}}
\,\,\,\,\,\,\,\,\,\,\,\,\,\,\,\,\,\,\,
\widetilde W^n=W^{n-1}.\cr}
\eqn\cietseis
$$
It is easy to check that $\widetilde G^n$, $\widetilde Q^n$
and $\widetilde V^n$ are the Laurent modes of
 dimension-${3\over 2}$ primary fields, whereas $\widetilde W^n$
corresponds to a dimension-2 Virasoro primary. Thus the
deformation \ciettres\ changes the dimensions of the operators
 appearing in our algebra. Moreover we can find a linear
combination of the dimension-${3\over 2}$ fields mentioned above
such  that behaves as supercurrent. Indeed if we define
$$
\widetilde {\cal G}^n=\widetilde Q^n +\widetilde G^n -{1\over
2}\widetilde V^n,
\eqn\cietsiete
$$
we can immediately check using the brackets of  Appendix A
that
$$
[\widetilde {\cal G}^n,\widetilde {\cal G}^m]=
2\widetilde T^{n+m}+d(n^2-{1\over 4})\delta_{m+n,0}
,\eqn\cietocho
$$
which corresponds to the anticommutator among the modes of the
supercurrent of an $N=1$ superconformal algebra. It is easy to
verify that the other fields appearing in the topological
algebra can be accommodated into $N=1$ supermultiplets. Recall
that such  multiplets are composed by two fields $(A,B)$ in
such a way that they are connected by the supercurrent. In
fact if $\Delta _A$ and $\Delta _B$ are  the conformal
dimensions of the fields $A$ and $B$ ($\Delta _B =\Delta _A
+{1\over 2}$) one has
$$
\eqalign{
[\widetilde {\cal G}^n, A^m]=&B^{n+m}\cr
[\widetilde {\cal G}^n, B^m]=&((2\Delta_A-1)n-m)A^{n+m}.\cr}
\eqn\cietnueve
$$
Two of such supermultiplets can be formed. First of all, if we
define
$$
S=-\widetilde Q+\widetilde G-{3\over 2}\widetilde V,
\eqn\ciecuarenta
$$
it can be checked after a short calculation that $S$ is the
supersymmetric companion of the $U(1)$ current $R$, \ie, $(R,S)$
behave as in eq. \cietnueve\ with $\Delta_A=1$.  The two
remaining fields $\widetilde V$ and $\widetilde W$ are the
components of the multiplet  $ (\widetilde V,\widetilde W)$
which satisfy eq.\ciecuarenta\ with conformal weight ${3\over
2}$.

\chapter {The $GL(N,N)$ Wess-Zumino-Witten model and free
fields}

In this section we shall study a two-dimensional field
theory  possessing a $gl(N,N)$ current algebra. This
model is nothing but a Wess-Zumino-Witten (WZW) model for
the $GL(N,N)$ supergroup. Our basic variable will  now be
a function $g$ taking values in $GL(N,N)$. As  is
well known, in a vicinity of the unit, $g$ can be
represented locally as the exponential of an element of
the $gl(N,N)$ algebra. Let us introduce complex
coordinates for our two-dimensional base manifold
according to the conventions:
$$
\eqalign{
z=x+iy
\,\,\,\,\,\,\,\,\,\,\,\,
\partial ={1\over 2}(\partial _x-i\partial_y)\cr
\bar z=x-iy
\,\,\,\,\,\,\,\,\,\,\,\,
\bar \partial ={1\over 2}(\partial
_x+i\partial_y),\cr}
\eqn\ffuno
$$
where $x$ and $y$ are arbitrary real coordinates. The action
for the WZW model is given by [\WZW]
$$
\eqalign{
S[g]=k\Gamma [g]=&
{k\over 2\pi}\int d^2x
Str(g^{-1}\partial g g^{-1}{\bar \partial} g ) +\cr
+&{ik\over 12\pi}\int d^3x \epsilon^{\mu \nu \rho}
Str(g^{-1}\partial_{\mu} g g^{-1}\partial_{\nu} g
g^{-1}\partial_{\rho} g ),\cr}
\eqn\ffdos
$$
where $k$ is a c-number constant (the level
of the current algebra) and, as  is well known, the
three-dimensional integral appearing in the right-hand
side of \ffdos\ is taken over a manifold whose boundary
is our two-dimensional base manifold. The functional
$\Gamma$ defined in \ffdos\ satisfies the remarkable
Polyakov-Wiegmann (PW) cocycle property
\REF\poly{A.M. Polyakov and P.W. Wiegmann
\journal\pl&B(83)121.}[\poly]:
$$
\Gamma[gh]=\Gamma[g]+\Gamma[h]+
{1\over \pi}\int d^2x
Str(g^{-1}{\bar \partial} g \partial h h^{-1}).
\eqn\fftres
$$
This equation can be demonstrated by a direct
calculation. The only difference between the proof of the
PW property for groups and supergroups is that, in the
latter case, we have to deal with supertraces instead of
ordinary traces. As the supertrace of elements of
$GL(N,N)$  satisfies the cyclic property, eq. \fftres\
can be obtained following the same steps as in the case of
ordinary groups.

One of the main advantages of working
with an explicit representation like the one based on the
action \ffdos\ is that one can convert the model into a
theory of free fields. The standard procedure to achieve
this objective has been described in ref [\gera]. The
first step consists in using a gaussian representation of
the (super)group variable. Any element of $GL(N,N)$ can
be decomposed as
$$
g=g_{_L} g_{_D} g_{_U}=
\pmatrix{1&0\cr \lambda&1\cr}
\pmatrix{\alpha&0\cr 0&\beta\cr}
\pmatrix{1&\chi\cr 0&1\cr},
\eqn\ffcuatro
$$
where we have used an $N\times N$ block notation in which
$\alpha$ and $\beta$ ($\chi$ and $\lambda$) are $N\times
N$ bosonic (respectively fermionic) matrices. Using the
PW property (eq. \fftres) we can readily write the action
$S$ in terms of the variables introduced in \ffcuatro:
$$
S[g]=k\gamma [\alpha]-k\gamma [\beta]-
{k\over \pi}\int d^2x
Tr(\alpha\partial\chi\beta^{-1}\bar\partial\lambda),
\eqn\ffcinco
$$
where $\gamma$ is the general WZW functional for the
$gl(N)$ group, whose explicit expression can be obtained
by replacing supertrace by trace in eq. \ffdos. The currents
implementing the $gl(N,N)$ Kac-Moody symmetry are given by
Witten's bosonization formula. Concentrating only on the
holomorphic part of the algebra we have:
$$
J=-k\partial g g^{-1}.
\eqn\ffseis
$$

The different contravariant components of the current can
be read from the matrix elements of $J$ as follows (see
eq. \vsiete):
$$
\eqalign{
J=&\pmatrix{J^{ab}&J^{a, b+N}\cr
J^{a+N,b}&J^{a+N, b+N}\cr}\cr}.
\eqn\ffsiete
$$
Substituting the gaussian decomposition \ffcuatro\ in eq.
\ffseis\ we obtain:
$$
\eqalign{
J^{ab}=&\,\,\,k(\alpha\partial \chi\beta^{-1}\lambda-
\partial\alpha\alpha^{-1})^{ab}\cr
J^{a, b+N}=&\,\,-k(\alpha\partial\chi\beta^{-1})^{ab}\cr
J^{a+N,b}=&\,\,\,k(\lambda\alpha\partial\chi\beta^{-1}\lambda-
\lambda\partial\alpha\alpha^{-1}+\partial\beta\beta^{-1}\lambda
-\partial\lambda)^{ab}\cr
J^{a+N,b+N}=&
\,\,-k(\lambda\alpha\partial\chi\beta^{-1}+
\partial\beta\beta^{-1})^{ab}.\cr}
\eqn\ffocho
$$
The action \ffcinco\ and the current components \ffocho\
are greatly simplified in terms of the fermionic
variables
$$
\eqalign{
\xi_{ab}=&\lambda ^{ba}\cr
\eta_{ab} =& k(\alpha\partial\chi\beta^{-1})^{ba}.\cr}
\eqn\ffnueve
$$
In fact the fermionic part of the action $S$ takes a
free-field form
$$
S[g]=k\gamma [\alpha]-k\gamma [\beta]-
{1\over \pi}\int d^2x Tr(\eta\bar\partial\xi).
\eqn\ffdiez
$$
Notice that the operator product expansion (OPE) between
$\eta$ and $\xi$ that follows from \ffdiez\ is simply
$$
\eta_{ab}(z)\xi_{cd}(w)=-{\delta_{ad}\delta_{bc}
\over z-w}.
\eqn\ffonce
$$
The bosonic $gl(N)$ currents corresponding to the
$\alpha$ and $\beta$ variables appearing in \ffdiez\ are
$$
j^{(\alpha)}_{ab}=-k(\partial\alpha\alpha^{-1})^{ba}
\,\,\,\,\,\,\,\,\,\,\,\,\,\,\,
j^{(\beta)}_{ab}=k(\partial\beta\beta^{-1})^{ba}.
\eqn\ffdoce
$$
In terms of the free fermionic fields $\eta$, $\xi$ and
the currents \ffdoce, the different components of $J$
 simplify greatly. First of all, using eqs. \vocho\ and
\cidos, we can write
$$
\eqalign{
K_{ab}=&J_{ab}=J^{ba}\cr
L_{ab}=&J_{a+N,b+N}=-J^{b+N,a+N}\cr
\Psi_{ab}=&J_{a+N,b}=J^{b,a+N}\cr
\Lambda_{ab}=&J_{a,b+N}=-J^{b+N,a},\cr}
\eqn\fftrece
$$
and, taking  eqs. \ffocho,\ffnueve\ and
\ffdoce\ into account, one gets:
$$
\eqalign{
K_{ab}=&-(\xi\eta)_{ab}+j^{(\alpha)}_{ab}\cr
L_{ab}=&-(\eta\xi)_{ab}+j^{(\beta)}_{ab}\cr
\Psi_{ab}=&-\eta_{ab}\cr
\Lambda_{ab}=&\,(\xi\eta\xi)_{ab}-(j^{(\alpha)}\xi)_{ab}-
(\xi j^{(\beta)})_{ab}+k\partial\xi_{ab}.\cr}
\eqn\ffcatorce
$$
Up to now, we have treated our fields as classical objects.
In a quantum theory, however, a change of variables as the one
written down in eq. \ffnueve\ should be accompanied by
the corresponding Jacobian, which is a functional
determinant that takes  the change in the
measure of the corresponding path integral into account.
These quantum
corrections can introduce additional terms in the naive
effective action \ffdiez, which can change the
propagators and couplings of the theory and, therefore, the
corresponding OPE's between the different operators. On the
other hand [\gera] the free-field expression of the
currents can also be affected by these quantum
corrections, since extra terms can originate from the
regularization of singular operator products in eq.
\ffocho. We will not attempt to compute this functional
determinant. Instead we are going to check whether or not
the currents \ffcatorce\ satisfy the OPE's of the $gl(N,N)$
affine algebra at level $k$. We shall see that this is
not the case if the OPE's dictated from the classical
action \ffdiez\ are used. Therefore, we must allow for a slight
modification of these basic OPE's in order to have a
representation of the $gl(N,N)$ Kac-Moody algebra. For
example, the bosonic $gl(N)$ currents $j^{(\alpha)}$ and
$j^{(\beta)}$ must satisfy the OPE's
$$
\eqalign{
j^{(\alpha)}_{ab}(z)j^{(\alpha)}_{cd}(w)=&
{k_\alpha\delta_{ad}\delta_{bc}\over (z-w)^2}+
{1\over z-w}[\delta_{bc}j^{(\alpha)}_{ad}(w)-
\delta_{ad}j^{(\alpha)}_{cb}(w)]\cr
j^{(\beta)}_{ab}(z)j^{(\beta)}_{cd}(w)=&
{k_\beta\delta_{ad}\delta_{bc}\over (z-w)^2}+
{1\over z-w}[\delta_{bc}j^{(\beta)}_{ad}(w)-
\delta_{ad}j^{(\beta)}_{cb}(w)],\cr}
\eqn\ffquince
$$
where we have allowed for a finite renormalization of the
levels. These constants $k_\alpha$ and $k_\beta$ can be
determined from the products of two bosonic
$gl(N,N)$ currents. Performing a direct calculation using
the free-field representation \ffcatorce\ and the
operator products \ffonce\ and \ffquince\ we get:
$$
\eqalign{
K_{ab}(z)K_{cd}(w)=&
{(k_\alpha+N)\delta_{ad}\delta_{bc}\over (z-w)^2}+
{1\over z-w}[\delta_{bc}K_{ad}(w)-
\delta_{ad}K_{cb}(w)]\cr
L_{ab}(z)L_{cd}(w)=&
{(k_\beta+N)\delta_{ad}\delta_{bc}\over (z-w)^2}+
{1\over z-w}[\delta_{bc}L_{ad}(w)-
\delta_{ad}L_{cb}(w)].\cr}
\eqn\ffdseis
$$
Comparing the right-hand side of \ffdseis\ with eq.
\citres\ we can determine the unknown constants $k_\alpha$
and $k_\beta$:
$$
k_\alpha=k-N
\,\,\,\,\,\,\,\,\,\,\,\,\,\,\,\,\,
k_\beta=-k-N.
\eqn\ffdsiete
$$
Notice that the $gl(N)$ levels of $j^{(\alpha)}$ and
$j^{(\beta)}$ are renormalized by the same quantity ($N$)
with respect to the values ($k$ and $-k$ respectively)
dictated by the classical action \ffdiez. At the
classical level $j^{(\alpha)}$ and
$j^{(\beta)}$ are uncoupled. However it is natural to
suppose that the jacobian of the change of variables
\ffnueve\ could induce terms  that couple these two currents
 in the quantum effective action.
Actually this coupling is needed  in order to reproduce the
$gl(N,N)$ current algebra with our our free-field representation.
If we compute the OPE of $K$ and $L$ we get
$$
K_{ab}(z)L_{cd}(w)=
-{\delta_{ab}\delta_{cd}\over (z-w)^2}+
j^{(\alpha)}_{ab}(z)j^{(\beta)}_{cd}(w),
\eqn\ffdocho
$$
and therefore in order to get a vanishing result (as in
\citres) we need to impose
$$
j^{(\alpha)}_{ab}(z)j^{(\beta)}_{cd}(w)=
{\delta_{ab}\delta_{cd}\over (z-w)^2}.
\eqn\ffdnueve
$$
Notice the different index structure of the double pole
in eqs. \ffdnueve\ and \ffquince. In this last equation,
the coefficient of the ${1\over (z-w)^2}$ term is
proportional to the $gl(N)$ Cartan-Killing form, whereas
in \ffdnueve\ the coupling only affects the abelian
subalgebra of $gl(N)$ (see below). We can now compute the
singular terms in the product expansion of fermionic and
bosonic currents with the result
$$
\eqalign{
\Psi_{ab}(z) K_{cd}(w)=&{\delta_{bc}\over
z-w}\Psi_{ad}(w)\cr
\Psi_{ab}(z)L_{cd}(w)=
&-{\delta_{ad}\over z-w}\Psi_{cb}(w)\cr
\Lambda_{ab}(z) K_{cd}(w)=&-{\delta_{ad}\over
z-w}\Lambda_{cb}(w)\cr
\Lambda_{ab}(z) L_{cd}(w)=&{\delta_{bc}\over
z-w}\Lambda_{ad}(w).\cr}
\eqn\ffveinte
$$
Eq. \ffveinte\ is in agreement with the brackets listed
in eq. \cicinco. The products  it remains to check are
those in which two fermionic currents are multiplied.
After some calculation we get
$$
\eqalign{
\Lambda_{ab}(z)\Lambda_{cd}(w)=&0\cr
\Psi_{ab}(z)\Psi_{cd}(w)=&0\cr
\Psi_{ab}(z)\Lambda_{cd}(w)=&
{k\delta_{bc}\delta_{ad}\over (z-w)^2}
-{1\over
z-w}[\delta_{ad}K_{cb}(w)+\delta_{bc}L_{ad}(w)].\cr}
\eqn\ffvuno
$$
When we compare the $\Psi\Lambda$ product in \ffvuno\ with
the corresponding anticommutator (eq. \cicuatro), we
notice that the right-hand side of the result obtained in
\ffvuno\ differs in  sign from the result we should have
found according to eq. \cicuatro. This problem can be
arranged by reversing the sign of $\Psi$ in our free-field
representation. Notice that none of  eqs. \ffveinte\
is altered by this change. Therefore we replace the
third equation in \ffcatorce\ by
$$
\Psi_{ab}=\eta_{ab}.
\eqn\ffvdos
$$
Once we have represented the currents by free fields,
we can obtain the expression of the different
operators appearing in the topological algebra. Let us
begin by considering the energy-momentum tensor. First of
all we denote the
Sugawara bilinears for the  $j^{(\alpha)}$ and
$j^{(\beta)}$ currents  by $T_1^{(\alpha)}$ and $T_1^{(\beta)}$
respectively:
$$
\eqalign{
T_1^{(\alpha)}=&\sum_{a,b}
:j^{(\alpha)}_{ab}j^{(\alpha)}_{ba}:\cr
T_1^{(\beta)}=&\sum_{a,b}
:j^{(\beta)}_{ab}j^{(\beta)}_{ba}:.\cr}
\eqn\ffvtres
$$
The operators $T_1$ and $T_2$ obtained with the first and
second Casimirs of $gl(N,N)$ are represented as
$$
\eqalign{
T_1=&T_1^{(\alpha)}-T_1^{(\beta)}-
N\sum_a(\partial j^{(\alpha)}_{aa}
+\partial j^{(\beta)}_{aa}) +2k\sum_a :\eta_{ab}\partial
\xi_{ba}: \cr
T_2=&\sum_{a,b} :(j^{(\alpha)}_{aa}+ j^{(\beta)}_{aa})
 (j^{(\alpha)}_{bb}+ j^{(\beta)}_{bb}):.\cr}
\eqn\ffvcuatro
$$
Adopting a trace notation, we can now write down the
expression of $T$ as
$$
\eqalign{
T=:Tr(\eta\partial \xi):+
&{1\over 2k}[T_1^{(\alpha)}-T_1^{(\beta)}-
NTr(\partial j^{(\alpha)}+ \partial j^{(\beta)})]\cr
+&{1\over 2k^2}[:Tr(j^{(\alpha)}+ j^{(\beta)})
Tr(j^{(\alpha)}+ j^{(\beta)}):].\cr}
\eqn\ffvcuatroi
$$
As a check one may verify that the currents $J_{AB}$
represented as in eqs. \ffcatorce\ and \ffvdos\ are primary
with respect to the operator \ffvcuatroi. In this
calculation one must use the singular product expansion
of $T_1^{(\alpha)}$ and $T_1^{(\beta)}$ with  $j^{(\alpha)}$ and
$j^{(\beta)}$. These can be obtained by using eqs.
\ffquince\ and \ffdnueve. One has
$$
\eqalign{
T_1^{(\alpha)}(z)j^{(\alpha)}_{ab}(w)=&
{2(k_\alpha+N)\over (z-w)^2}j^{(\alpha)}_{ab}(w)+
{2(k_\alpha+N)\over z-w}\partial
j^{(\alpha)}_{ab}(w)-\cr
&-{2\delta_{ab}\over (z-w)^2}\sum_c j^{(\alpha)}_{cc}(w)-
{2\delta_{ab}\over z-w}\sum_c
\partial j^{(\alpha)}_{cc}(w).\cr}
\eqn\ffvcinco
$$
This expression can be obtained in a way similar to the
one employed to prove eq. \cdos. Moreover, due to
the coupling \ffdnueve, the product
$T_1^{(\alpha)}j^{(\beta)}$ contains singular terms. It
can be easily proved that
$$
T_1^{(\alpha)}(z)j^{(\beta)}_{ab}(w)=
{2\delta_{ab}\over (z-w)^2}\sum_c j^{(\alpha)}_{cc}(w)+
{2\delta_{ab}\over z-w}\sum_c
\partial j^{(\alpha)}_{cc}(w).
\eqn\ffvseis
$$
Of course there are equations similar to \ffvcinco\ and
\ffvseis\ in which the labels $\alpha$ and $\beta$ are
interchanged. It is now straightforward to check that
the $J_{AB}$ currents satisfy
$$
T(z)J_{AB}(w)={1\over (z-w)^2}J_{AB}(w)+
{1\over (z-w)}\partial J_{AB}(w),
\eqn\ffvsiete
$$
which is the OPE equivalent to the bracket \cseis. Other
generators appearing in the topological algebra can be
equally computed:
$$
\eqalign{
G=& {1\over 2k}:Tr [\eta (j^{(\alpha)}-j^{(\beta)})] :
-{N\over 2k}Tr(\partial \eta )
+{1\over 2k^2}:Tr(j^{(\alpha)}+j^{(\beta)})
Tr(\eta):\cr
Q=&-: Tr[ \xi (j^{(\alpha)}+j^{(\beta)})
-\xi\eta\xi-k\partial \xi]:\cr
R=&{1\over 2}Tr(j^{(\alpha)}-j^{(\beta)})+:Tr(\eta\xi):\cr
W=&-{1\over 2k^2}:Tr[(j^{(\alpha)}+j^{(\beta)})\eta\eta]:
+{1\over 2k^2}:[Tr\partial \eta Tr \eta -NTr(\partial
\eta\eta)]:\cr
V=&-{1\over 6k^2} :Tr(\eta^3):.\cr}
\eqn\ffvocho
$$
Eqs. \ffvcuatroi\ and \ffvocho\ provide an explicit
representation of our topological algebra in terms of two
bosonic currents $j^{(\alpha)}$ and $j^{(\beta)}$ and two
fermionic fields $\eta$ and $\xi$. However, as we pointed
out above, the abelian components of   $j^{(\alpha)}$ and
$j^{(\beta)}$ are coupled (see eq. \ffdnueve). A free-field
representation of an algebra should be given in terms of a set
of uncoupled fields. As the coupling of the traceless
components of the currents vanishes, it will be convenient to
separate the $sl(N)$ parts of  $j^{(\alpha)}$ and $j^{(\beta)}$
from the remaining $U(1)$ contributions. This decomposition
will serve us to clarify the nature of the extended topological
algebra. Actually we show below that our algebra is intimately
linked to the non-abelian topological current algebra
 underlying our theory.

We split our currents as:
$$
\eqalign{
j_{ab}^{(\alpha)}=&I_{ab}^{(\alpha)}+\delta_{ab}
\sqrt {{N-k\over N}}\,\,j_1\cr
j_{ab}^{(\beta)}=&I_{ab}^{(\beta)}-\delta_{ab}
\sqrt {{N+k\over N}}\,\,j_2,\cr}
\eqn\ffvnueve
$$
where the $sl(N)$ currents $I^{(\alpha)}$ and
$I^{(\beta)}$ are traceless:
$$
\sum_a I_{aa}^{(\alpha)}=\sum_a I_{aa}^{(\beta)}=0.
\eqn\fftreinta
$$
The coefficients multiplying the abelian currents $j_1$
and $j_2$ have been chosen for later convenience. Using
eqs. \ffquince\ and \ffdnueve\ we can write the OPE's
 satisfied by our $sl(N)$ currents
$$
\eqalign{
I_{ab}^{(\alpha)}(z)I_{cd}^{(\alpha)}(w)=&
(k-N){\delta_{ad}\delta_{bc}-{1\over
N}\delta_{ab}\delta_{cd}\over (z-w)^2}+
{\delta_{bc}I_{ad}^{(\alpha)}(w)-
\delta_{ad}I_{cb}^{(\alpha)}(w)\over z-w}\cr
I_{ab}^{(\beta)}(z)I_{cd}^{(\beta)}(w)=&
-(k+N){\delta_{ad}\delta_{bc}-{1\over
N}\delta_{ab}\delta_{cd}\over (z-w)^2}
+{\delta_{bc}I_{ad}^{(\beta)}(w)-
\delta_{ad}I_{cb}^{(\beta)}(w)\over z-w}.\cr}
\eqn\fftuno
$$
The $U(1)$ currents have the following singular product
expansions:
$$
\eqalign{
j_1(z)j_1(w)=&j_2(z)j_2(w)=-{1\over (z-w)^2}\cr
j_1(z)j_2(w)=&-{N\over (N^2-K^2)^{1\over 2}}\,\,{1\over
(z-w)^2}.\cr}
\eqn\fftdos
$$
Other products of currents appearing in the decomposition
\ffvnueve\ are regular. The OPE's of eq. \fftdos\ can be
reproduced by representing $j_1$ and $j_2$ in terms of
two bosonic fields $\varphi$ and $\phi$. Let us assume
that these two fields obey the OPE
$$
\varphi(z)\phi(w)=-log(z-w).
\eqn\ffttres
$$
Then eq. \fftdos\ is satisfied by the following
combinations of $\partial \varphi$ and $\partial \phi$:
$$
\eqalign{
j_1={1\over (N-k)^{1\over 2}}\partial \varphi
+{(N-k)^{1\over2}\over 2}\partial \phi\cr
j_2={1\over (N+k)^{1\over 2}}\partial \varphi
+{(N+k)^{1\over2}\over 2}\partial \phi.\cr}
\eqn\fftcuatro
$$

Let us now perform a decomposition of the fermionic
fields $\eta$ and $\xi$ similar to the one adopted in the
bosonic sector of the theory:
$$
\eta_{ab}=\rho_{ab}+k{\delta_{ab}\over \sqrt{N}}\mu
\,\,\,\,\,\,\,\,\,\,\,\,\,\,\,\,\,\,
\xi_{ab}=\gamma_{ab}+{\delta_{ab}\over k\sqrt{N}}\zeta ,
\eqn\fftcinco
$$
where the $\gamma$ and $\rho$ fields are traceless, \ie :
$$
\sum_a\rho_{aa}=\sum_a\gamma_{aa}=0.
\eqn\fftseis
$$
The only non-vanishing OPE's are:
$$
\rho_{ab}(z)\gamma_{cd}(w)=
-{\delta_{ad}\delta_{bc}-{1\over N}\delta_{ab}\delta_{cd}
\over z-w}
\,\,\,\,\,\,\,\,\,\,\,\,\,\,\,\,\,\,
\mu(z)\zeta(w)=-{1\over z-w}.
\eqn\fftsiete
$$
Substituting eqs. \ffvnueve, \fftcuatro\ and \fftcinco\ into the
expression of the energy-momentum tensor of the theory
(eq. \ffvcuatroi) we obtain $T$ as the sum of an $sl(N)$
contribution and a $U(1)$ part:
$$
T=T_{sl(N)}+T_{U(1)}.
\eqn\fftocho
$$
where
$$
\eqalign{
T_{sl(N)}=&{1\over 2k}[:Tr(I^{(\alpha)}I^{(\alpha)}):
-:Tr(I^{(\beta)}I^{(\beta)}):]+ :Tr(\rho\partial\gamma):\cr
T_{U(1)}=&-:\partial\phi\partial\varphi:+{N^{3\over
2}\over 2}\,\,\partial^2\phi +:\mu\partial\zeta:.\cr}
\eqn\fftnueve
$$
This additivity of the $sl(N)$ and $U(1)$ components is
satisfied by all the generators of the topological
algebra.

It is interesting to notice that both operators
$T_{sl(N)}$ and $T_{U(1)}$ have a vanishing Virasoro
anomaly. This means that the $sl(N)$ and $U(1)$ parts of
our theory constitute separate  topological conformal
field theories. In fact they are the irreducible components
of our topological theory. There is however a fundamental
difference between these two components: while the
non-abelian part non-trivially satisfies  the extended
topological algebra, the algebra of the $U(1)$ component
closes without introducing spin-three operators. In this
sense we can say that the non-abelian part is responsible
for the extension of the algebra. Another important point
is that the dimension $d$ of these two algebras is not
zero, although their sum vanishes. Actually the dimension
of the non-abelian part exactly equals the dimension of
the $SL(N)$ group manifold. This fact suggests that our
theory can be regarded as a topological sigma model for
the group manifold $SL(N)$ (\ie\ as a topological version
of the $SL(N)$ WZW model) together with a compensating
abelian component.

Let us first study the $U(1)$ component of our theory.
The generators of the topological algebra have the
following form:
$$
\eqalign{
G=&:\mu\partial\varphi:-{N^{3\over 2}\,\,\over 2}\partial
\mu\cr
Q=&:\zeta\partial\phi:+\sqrt{N}\partial\zeta\cr
R=&\sqrt{N}(\partial \varphi +{N\over 2}\partial \phi)
+:\mu\zeta:.\cr}
\eqn\ffcuarenta
$$
The $U(1)$ contribution to $V$ and $W$ vanishes and the
parameter $d$  appearing in the algebra closed by the
operators \ffcuarenta\ is $1-N^2$. In order to see how
the BRST symmetry relates our fields, let us compute the
transformation generated on them by $Q$ . After a simple
calculation we get
$$
\eqalign{
\delta\varphi=&-\zeta
\,\,\,\,\,\,\,\,\,\,\,\,\,\,\,\,\,\,
\delta \phi=0\cr
\delta \mu=&-\partial\phi
\,\,\,\,\,\,\,\,\,\,\,\,\,\,
\delta \zeta=0,\cr}
\eqn\ffcuno
$$
which implies that $\zeta$ is the BRST copy of $\varphi$ and
$\mu$ is the partner of the abelian current $\partial
\phi$. In the $gl(1,1)$ case, this $U(1)$ contribution is the
only one. Its topological dimension in this case vanishes. As
explained in the previous section, by twisting the theory as in
eq. \cievsiete\ and \cievocho, we can get a non-zero $d$ value.
It is easy to verify that, in so doing,  we get the standard
representation [\li] of the twisted $N=2$ superconformal
algebra.

In order to analyse the non-abelian component of our
theory, let us adopt a more standard notation for the
$sl(N)$ Lie algebra. Let  $T^i$
($i=1,\cdots,N^2-~1$) be a set of generators of $sl(N)$
chosen in such a way that   $Tr(T^iT^j)=\delta^{ij}$. They
satisfy commutation relations $[T^i,T^j]=f^{ijk}T^k$, where
$f^{ijk}$ are the (totally antisymmetric) structure constants.
If $A$ is an arbitrary element of the Lie algebra, its
components $A_{ab}$ with respect to the $E_{ab}$ matrices and
$A^i$ with respect to the generators $T^i$ are related by the
expressions $$
A^i=\sum_{ab}A_{ab}(T^i)_{ab}
\,\,\,\,\,\,\,\,\,\,\,\,\,\,\,\,\,\,
A_{ab}=\sum_iA^i(T^i)_{ba}.
\eqn\ffcdos
$$
Using these equations, the $sl(N)$ contribution to the
operators of eq. \ffvocho\ can be written as
$$
\eqalign{
G=&{1\over 2k}:\rho^i(I^{{(\alpha)}\,^i}-
I^{{(\beta)}\,^i}):\cr
Q=&-:\gamma^i(I^{{(\alpha)}\,^i}+
I^{{(\beta)}\,^i}+{1\over 2}f^{ijk}\gamma^{j}\rho^k):\cr
R=&:\rho^i\gamma^i:\cr
W=&{1\over 4k^2}:f^{ijk}(I^{{(\alpha)}\,^i}+
I^{{(\beta)}\,^i})\rho^j\rho^k:-
{N\over 2k^2}:\partial\rho^i\rho^i:\cr
V=&{1\over 12 k^2}:f^{ijk}\rho^i\rho^j\rho^k:,\cr}
\eqn\ffctres
$$
where we sum over repeated indices.
As we mentioned above, the dimension $d$ appearing in the
algebra satisfied by the operators of eq. \ffctres\ is
$$
d_{sl(N)}=N^2-1.
$$
Let us denote by ${\cal I}$ the $sl(N)$ contribution to the
current ${\cal J}$. Its components along the $T^i$
generators are given by
$$
{\cal I}^i=I^{{(\alpha)}\,^i}+
I^{{(\beta)}\,^i}+:f^{ijk}\gamma^{j}\rho^k:.
\eqn\ffccinco
$$
It is instructive to write the BRST transformation of the
non-abelian fields:
$$
\eqalign{
\delta I^{{(\alpha)}\,^i}=&f^{ijk}\gamma^j I^{{(\alpha)}\,^k}
 +(N-k)\partial \gamma^i\cr
\delta I^{{(\beta)}\,^i}=&f^{ijk}\gamma^j I^{{(\beta)}\,^k}
 +(N+k)\partial \gamma^i\cr
\delta \gamma^i=&{1\over 2}f^{ijk}\gamma^j\gamma^k\cr
\delta \rho^i=&I^{{(\alpha)}\,^i}+I^{{(\beta)}\,^i}+
f^{ijk}\gamma^j\rho^k,\cr}
\eqn\ffcseis
$$
The BRST transformation of our
$gl(N)$ fields has, in fact, the standard form of a
non-abelian BRST symmetry. This transformation law implies
that the field $\rho$ is the partner of ${\cal I}$. They
close an $sl(N)$ algebra without central extension:
$$
\eqalign{
{\cal I}^i(z){\cal I}^j(w)=&f^{ijk}{{\cal I}^k(w)\over z-w}\cr
\rho^i(z){\cal I}^j(w)=&f^{ijk}{\rho ^k(w)\over z-w}.\cr}
\eqn\ffcsiete
$$
Notice that, although the level $k$ does not appear in the
current algebra, it shows up in the BRST variations of the
currents. In fact the last two terms in the first two
equations \ffcseis\  are proportional to $k_{\alpha}$ and
$k_{\beta}$, which are the central extensions appearing in
\fftuno.

The form of the energy-momentum tensor $T$, its BRST partner
$G$ and the two dimension-one currents $Q$ and $R$, is
identical to the one found in refs. [\yank,\aharo,\hu] for the
$sl(N)/sl(N)$ topological coset models. This means that the
extended topological symmetry we have obtained is realized in
the $G/G$ theories. Moreover, the deformations \cievsiete\ and
\cievocho\ found in the previous section , when restricted to the
$sl(N)$ component, take the form:
$$
\eqalign{
T\rightarrow &T+\sum_i\alpha^i\partial {\cal I}^i\cr
G\rightarrow &G+\sum_i\alpha^i\partial \rho^i\cr
R\rightarrow &R+\sum_i\alpha^i{\cal I}^i.\cr}
\eqn\ffcocho
$$
The parameter $d_{sl(N)}$ is not changed from its value
$N^2-1$ under this deformation. This  can be easily
verified by explicit calculation but, in fact, it is clear
from \cietreinta\ that only the $U(1)$ sector of our theory
changes its parameter $d$ under this transformation.
Remarkably, when the $\alpha^i$ are taken equal to one for the
 generators of the $sl(N)$ Cartan subalgebra and zero otherwise,
it has been noticed in refs. [\aharo,\hu] that the deformed
$sl(N)/sl(N)$ model is equivalent to a system of $(p,q)$ $W_N$
minimal matter coupled to $W_N$ gravity (plus some extra
topological sectors). With our notations, this equivalence is
valid when $k$ is equal to ${p\over 2q}$, with $p,q$ integers.
Actually this is not surprising in view of the interpretation
of the $G/G$ models as the topological analogue of the WZW
model. A similar result was obtained in refs.
\REF\kpz{A.M. Polyakov \journal\mpl&A2(87)893; V. Knizhnik,
A.M. Polyakov and A.B. Zamolodchikov \journal\mpl&A3(88)819.}
\REF\BO{M. Bershadski and H. Ooguri \journal\cmp&126(89)49.}
[\kpz,\BO]
 using the quantum Hamiltonian reduction of the WZW model. It
is interesting to point out that, since the dimension-three
generators $V$ and $W$ are not altered under the
transformation \ffcocho, the  algebra of the
modified theory retains its extended character after the
deformation has been performed. The fact that our extended
algebra is present in the minimal topological matter systems
leads us to conjecture that, in fact, we are dealing with the
basic topological algebra.

\chapter{Discussion and outlook}

The theories with a current algebra symmetry are the basic
building blocks from which all known rational conformal field
theories can be constructed. It might be that a similar
statement could apply to the two-dimensional topological
conformal field theories. In this paper we have constructed a
model  possessing both topological and current algebra
symmetries. The currents that close with the topological
algebra are of a special type: they form a topological
multiplet, composed by a bosonic current and its BRST partner,
and they obey an algebra without central extension. A c-number
anomaly appears, however, when the BRST symmetry acts on the
currents. Contrary to what happens with the Virasoro algebra,
in order to incorporate a current algebra with a non-abelian
topological symmetry, one has to extend the topological
algebra by including spin-three generators. Although we do not
have a general proof of this statement, this conclusion is
likely to hold for any BRST symmetry acting on the currents
in a non-abelian way (\ie\ as an odd rotation with central
extension).

An important point we have not considered here is the
determination of the spectrum of physical states of our theory.
Closely related to this question is the analysis of the
topological invariant observables of the model. In view of
the connection found in section 5 with the $G/G$ coset models,
it is clear that one could invoke the results of refs.
[\aharo,\hu], in which the cohomology of the Fock space of the
topological $G/G$ theory has been studied.

One of our motivations to study a theory with an affine
$gl(N,N)$ symmetry was the relation found in ref. [\RS ]
between the $gl(1,1)$ theory and the Alexander-Conway knot
polynomial. It would be interesting to analyse the implications
of the topological symmetry found here
in the study of knot invariants. In particular, for $N>1$, our
analysis could lead to formulate a non-abelian generalization
of the Alexander-Conway knot polynomial. It is interesting to
notice in this respect that, as  was stressed in ref.
[\yank], the $G/G$ coset models are the two-dimensional
analogues of the three-dimensional Chern-Simons theory for the
gauge group G. This fact is a clue that may help to understand
the precise relation between the Alexander-Conway and the Jones
polynomials, a problem that remains open in knot theory.

Finally, let us point out that, although we have restricted
ourselves to two-dimensional space-times, we could try to use
a $gl(N,N)$ symmetry as a tool to generate topological field
theories in any number of dimensions. In the particular case
of four dimensions, it would be interesting to investigate if
one can generate models of topological matter and, in that
case,  study their coupling to gravity. We expect to report on
this and related issues in a near future.

\ack

We thank J.M.F. Labastida for useful discussions and
encouragement and J.~S\'anchez Guill\'en  for a critical reading
of the manuscript.  We are grateful to J.~Mas for pointing
out to us an error that slipped into a preliminary version of this
paper. Thanks are also given to  M. Alvarez and P.M. Llatas. This
work was supported in part by DGICYT under grant PB90-0772, and by
CICYT under grants  AEN90-0035 and AEN93-0729.
\endpage

\Appendix A
Below we collect the brackets of the extended topological
algebra for an arbitrary dimension $d$:

$$
\eqalign{
[T^n,T^m]=&(n-m)T^{n+m}\cr
[Q^n,Q^m]=&0\cr
[Q^n,G^m]=&T^{n+m}+nR^{n+m}+{d\over 2}m(m+1)\delta_{n+m,0}\cr
[T^n,R^m]=&-mR^{n+m}-{d\over 2}n(n+1)\delta_{n+m,0}\cr
[T^n,Q^m]=&-mQ^{n+m}\cr
[T^n,G^m]=&(n-m)G_{n+m}\cr
[R^n,R^m]=&dn\delta_{n+m,0}\cr
[R^n,Q^m]=&Q^{n+m}\cr
[R^n,G^m]=&-G^{n+m}\cr
[G^n,G^m]=&W^{n+m}\cr
[Q^n,W^m]=&0\cr
[R^n,W^m]=&-2W^{n+m}\cr
[T^n,W^m]=&(2n-m)W^{n+m}\cr
[G^n,W^m]=&(2n-m)V^{n+m}\cr
[Q^n,V^m]=&W^{n+m}\cr
[R^n,V^m]=&-3V^{n+m}\cr
[T^n,V^m]=&(2n-m)V^{n+m}\cr
[G^n,V^m]=&0\cr
[W^n,W^m]=&0\cr
[V^n,W^m]=&0\cr
[V^n,V^m]=&0.\cr}
\eqn\auno
$$

\Appendix B

In this appendix we shall show how a topological current
algebra of the type discussed in section 4 appears
in another context [\yos]. The model we shall analyse is
the zero-level $sl(2)$ WZW model. As the central charge
for the $sl(2)_k$ current algebra is $c={3k\over k+2}$,
it is clear that we shall be dealing with a conformal
field theory with vanishing Virasoro anomaly,
which we would expect  to be topological.  At the
topological point $k=0$, the $sl(2)$ currents close a
bosonic algebra without central extension, and there exists a
BRST symmetry making the energy-momentum tensor and the
currents cohomologically exact.

Let us start our analysis by recalling the free-field
representation of the $sl(2)$ affine algebra. It can be
obtained [\gera] by means of a gaussian decomposition of
the $SL(2)$ group similar to the one performed in
section 5. Let be $\phi$ a real scalar field and $\beta$
and $\gamma$ a bosonic system of $1$-- and
$0$--differentials obeying the OPE's
$$
\eqalign{
\phi (z) \phi(w)=-log(z-w)\cr
\beta (z)\gamma (w)=- {1\over z-w}.\cr}
\eqn\sluno
$$
The $sl(2)_k$ currents  are represented as
$$
\eqalign{
J_+=&\beta\cr
J_-=&-\beta\gamma^2-i\sqrt{2(k+2)}\gamma\partial\phi
-k\partial\gamma\cr
J_3=&\beta\gamma +i\sqrt{k+2\over 2}\partial \phi.\cr}
\eqn\sldos
$$
Using eq. \sldos\ it is straightforward to check that
they satisfy the OPE's
$$
\eqalign{
J_3(z)J_{\pm}(w)=\pm {1\over z-w}J_{\pm}(w)\cr
J_+(z)J_-(w)={k\over (z-w)^2}+{2\over
z-w}J_3(w).\cr}
\eqn\sltres
$$
The energy-momentum tensor of the theory can be witten in
terms of these free fields as
$$
\eqalign{
T=&{1\over 2(k+2)}:[J_+J_-+J_-J_++2(J_3)^2]:=\cr
=&-\beta\partial \gamma -{1\over 2}(\partial \phi)^2-
{i\over \sqrt{2(k+2)}}\partial^2\phi.\cr}
\eqn\slcuatro
$$
In order to see how the BRST symmetry appears at
the topological point, it is convenient to replace the
scalar field $\phi$ by a Grassmann $(b,c)$ system. These
new fields are related to  $\phi$  in the
following way:
$$
b=:e^{i\phi}:
\,\,\,\,\,\,\,\,\,\,\,\,\,
c=:e^{-i\phi}:.
\eqn\slcinco
$$
{}From the expansion \sluno, one gets
$$
b(z)c(w)={1\over z-w}.
\eqn\slseis
$$
On the other hand $T$ can be written in terms on $b$ and
$c$ as
$$
T=-\beta\partial \gamma-jb\partial c +(1-j)\partial b c
\eqn\slsiete
$$
where $j={1\over 2} +{1\over \sqrt{2(k+2)}}$ is the spin
of the field $b$, whereas $c$ has conformal weight
$1-j$. Taking into account that $:bc:=i\partial
\phi$, one can immediately express the currents in
terms of $b$ and $c$. Let us now put $k=0$ everywhere. The
$c$ field has  dimension zero  at this point and,
therefore, the fermionic nilpotent operator
$$
Q=:\beta c:
\eqn\slocho
$$
has dimension one. Let us see that this is the BRST
current we are looking for. First of all $Q$ induces the
following transformation on the fields:
$$
\eqalign{
\delta b=&\beta
\,\,\,\,\,\,\,\,\,\,\,\,\,\,\,
\delta \beta=0\cr
\delta \gamma=&-c
\,\,\,\,\,\,\,\,\,\,\,\,\,\,\,
\delta c=0.\cr}
\eqn\slnueve
$$

Now, putting $k=0$ in eq. \sldos, one
realizes that the $J^a$  operators are BRST-exact, \ie\
that there exist fermionic fields $\Psi^a$ such that
$J^a=\delta \Psi^a$. The explicit expression of these new
fields are
$$
\Psi_+=b
\,\,\,\,\,\,\,
\Psi_-=-b\gamma ^2
\,\,\,\,\,\,\,
\Psi_3=b\gamma.
\eqn\sldiez
$$
The $\Psi$'s are primary dimension-one operators with
respect to $T$. They close the
following algebra with the bosonic currents
$$
\Psi_a(z)J_b(w)={f^{abc}\over z-w}\Psi_c(w)
\,\,\,\,\,\,\,\,\,\,\,\,\,\,\,\,\,\,\,
\Psi_a(z)\Psi_b(w)=0,
\eqn\slonce
$$
which, together with eq. \sldos\ (for $k=0$), constitute what
we have called a topological current algebra. In \slonce\
$f^{abc}$ are the $sl(2)$ structure constants (see eq.
\sltres). The BRST partner of $T$, denoted as usual by
$G$, can be obtained as a Sugawara bilinear:
$$
G={1\over 4}:[\Psi_+J_-+\Psi_-J_++2\Psi_3J_3]:
=-:b\partial\gamma:.
\eqn\sldoce
$$
Notice that the global coefficient ${1\over 4}$ is the
same as in eq. \slcuatro\ for $k=0$. The operators $Q$, $T$
and $G$ close a (non-extended) topological algebra with
abelian current $R$ and dimension $d$ equal to
$$
R=:bc:
\,\,\,\,\,\,\,\,\,\,\,\,\,\,\,\,\,\,\,
d=1.
\eqn\sltrece
$$

It is interesting to obtain the OPE's of the BRST current
$Q$ and the $sl(2)$ currents. The singular terms in the
expansion of $Q(z)\Psi^a(w)$ are given by
$$
\eqalign{
Q(z)\Psi_+(w)=&{J_+(w)\over z-w}\cr
Q(z)\Psi_-(w)=&{2\over (z-w)^2}\gamma(w)+
{1\over z-w}J_-(w)\cr
Q(z)\Psi_3(w)=&{-1\over (z-w)^2}+
{1\over z-w}J_3(w).\cr}
\eqn\slcatorce
$$
It is worth  pointing out that
the central extensions appearing in eq. \slcatorce\ are
not $c$-numbers. In fact the same type of phenomenon
occurs when the products $Q(z)J^a(w)$ are computed:
$$
\eqalign{
Q(z)J_+(w)=&Q(z)J_3(w)=0\cr
Q(z)J_-(w)=&{2\over (z-w)^2}\,\,\,c(w).\cr}
\eqn\slquince
$$
Notice that the fields appearing in the double pole
singularity in eqs. \slcatorce\ and \slquince\ are the
zero-dimensional fields $\gamma$ and $c$. From \slnueve\
it follows that they belong to the same topological
multiplet. Other OPE's of the generators of the
topological algebra with the currents can also be
computed. For example, if we multiply $G$ by $J$ and
$\Psi$ we get
$$
\eqalign{
G(z)J_a(w)=&{\Psi_a(w)\over (z-w)^2}+
{\partial \Psi_a(w)\over z-w}\cr
G(z)\Psi_a(w)=&0.\cr}
\eqn\sldseis
$$
which mean that $J$ and $\Psi$ are also primary with
respect to $G$ in the sense discussed in section 4. The
behaviour of the currents under the $R$-symmetry is
determined by the expansions
$$
\eqalign{
R(z)J_+(w)=&0
\,\,\,\,\,\,\,\,\,\,
R(z)J_-(w)=-{2\over (z-w)^2}\,\,\,\gamma(w)\cr
R(z)J_3(w)=&{1\over (z-w)^2}
\,\,\,\,\,\,\,\,\,\,
R(z)\Psi_a(w)={\Psi_a(w)\over z-w}.\cr}
\eqn\sldsiete
$$
There is a fundamental difference between this system and the
one studied at the end of section 5. Although the currents $J$
and $\Psi$ close a non-abelian algebra, the BRST operator $Q$
does not act as a non-abelian symmetry on the currents. This
is consistent with the fact that the topological dimension of
this system is not equal to the dimension of $SL(2)$.

\endpage
\refout
\end